\documentclass[journal]{IEEEtran}

\usepackage{graphicx,psfrag,epsfig,epsf,latexsym,hhline,amsmath,amssymb,multirow}
\usepackage{pst-plot}
\usepackage{pstricks-add}
\usepackage{pifont}
\usepackage{graphicx}
\usepackage{amsthm}
\usepackage[linesnumbered,ruled,vlined]{algorithm2e}
\usepackage[noadjust]{cite}
\usepackage{array}
\usepackage{blindtext}
\usepackage{etoolbox}
\usepackage{comment}
\usepackage{epstopdf}
\usepackage{hyperref}
\usepackage{slashbox}
\usepackage{bbm}
\usepackage{arydshln}
\usepackage{lipsum}
\usepackage[flushleft]{threeparttable}

\newcommand\xqed[1]{%
  \leavevmode\unskip\penalty9999 \hbox{}\nobreak\hfill
  \quad\hbox{#1}}
\newcommand\demo{\xqed{$\blacksquare$}}

\SetCommentSty{mycommfont}

\SetKwInput{KwInput}{Input}                
\SetKwInput{KwOutput}{Output}              

\newcolumntype{M}[1]{>{\centering\arraybackslash}m{#1}}
\newcolumntype{P}[1]{>{\centering\arraybackslash}p{#1}}

\newcommand{\E}{\mathbb{E}}

\newcommand{\iid}{i.\@i.\@d.\ }

\theoremstyle{definition}

\newtheorem{example}{Example}
\newtheorem{theorem}{Theorem}
\newtheorem{lemma}{Lemma}

\newtheorem{remark}{Remark}
\newtheorem{corollary}{Corollary}

\begin{document}
\title{Sub-Block Rearranged Staircase Codes}
\author{Min Qiu~\IEEEmembership{Member,~IEEE,} and Jinhong Yuan~\IEEEmembership{Fellow,~IEEE}

\thanks{The work was supported in part by the Australian Research Council (ARC) Discovery Project under Grant DP220103596, and in part by the ARC Linkage Project under Grant LP200301482.}

\thanks{This work was presented in part at the 2022 IEEE Internal Symposium on Information Theory (ISIT) \cite{SRSCISIT22}.

The authors are with the School of Electrical Engineering and Telecommunications, University of New South Wales, Sydney, NSW, 2052 Australia (e-mail: min.qiu@unsw.edu.au; j.yuan@unsw.edu.au).

}%
}

\maketitle

\begin{abstract}
We propose a new family of spatially coupled product codes, called sub-block rearranged staircase (SR-staircase) codes. Each code block of SR-staircase codes is obtained by encoding rearranged preceding code blocks and new information blocks, where the rearrangement involves sub-blocks decomposition and transposition. The proposed codes can be constructed to have each code block size of $1/q$ to that of the conventional staircase codes while having the same rate and component codes, for any positive integer $q$. In this regard, we can use strong algebraic component codes to construct SR-staircase codes with a similar or the same code block size and rate as staircase codes with weak component codes. We characterize the decoding threshold of the proposed codes under iterative bounded distance decoding (iBDD) by using density evolution. We also derive the conditions under which they achieve a better decoding threshold than that of staircase codes. Further, we investigate the error floor performance by analyzing the contributing error patterns and their multiplicities. Both theoretical and simulation results show that the designed SR-staircase codes outperform staircase codes in terms of waterfall and error floor while the performance can be further improved by using a large coupling width.
\end{abstract}


\section{Introduction}
The explosive growth of data-hungry applications such as video streaming services and social networks has driven the development of high-speed optical networks. Forward error correction (FEC) codes are employed in optical communication systems to guarantee reliable data transmission. In particular, modern high-speed optical communication systems require FEC schemes: 1) to support throughput of 100 Gbit/s and beyond; 2) to have low power consumption; 3) to achieve a large coding gain close to the theoretical capacity limits at a target bit error rate (BER) of $10^{-15}$; and 4) to be adapted to the peculiarities of the optical channel \cite{6805157,optical_FEC_2020}.

A number of FEC codes that are popular for handling error correction in wireless communications have also been considered for optical communications \cite{6917198,optical_FEC_2020}. Among these FEC codes, low-density parity-check (LDPC) codes \cite{1057683} and spatially coupled LDPC codes \cite{782171} have gain much attention \cite{7340116} due to their provably close-to-capacity performance under belief propagation (BP) decoding \cite{5571910,5695130}. That said, the exchange of soft messages within their BP decoders significantly increases the internal data flow \cite{Smith12} as well as hardware and power cost for enabling high-resolution analog-to-digital conversion. An alternative solution is to resort to low-complexity hard decision decoding (HDD), which has significantly lower data flow \cite{Smith12} but suffer from some performance degradation. The FEC codes that are particularly suitable for high throughput HDD are product-like codes \cite{1057464} with Bose-Chaudhuri-Hocquengham (BCH) or Reed-Solomon component codes \cite{6917198}. HDD is performed iteratively by decoding the component codes using algebraic bounded distance decoding \cite{Richardson:2008:MCT:1795974}, which is referred to as iterative bounded distance decoding (iBDD) \cite{optical_FEC_2020}. Owing to the low-complexity decoding, product codes with iBDD have been adopted in various optical communications standards, e.g., \cite{G7951}.

Product-like codes continue to evolve today for achieving larger net coding gains. The authors in \cite{Smith12} applied the idea of spatial coupling to product codes with BCH component codes and constructed staircase codes. Remarkably, it was shown that their error performance only has a gap of 0.56 dB from the binary symmetric channel (BSC) capacity under iBDD and outperform existing FEC solutions in ITU-T G.975.1 \cite{G7951}. Another class of spatially coupled product codes called braided BCH codes were introduced in \cite{4957627}, which have comparable error performance to staircase codes. Both codes \cite{Smith12,4957627} can be considered as instances of spatially coupled generalized LDPC (GLDPC) ensembles \cite{1056404,767979} with BCH component codes as constraints. \cite{7954697} has proved that this class of spatially coupled GLDPC ensembles under iterative hard-decision decoding can approach capacity at high rates. A unified framework called zipper codes was recently proposed in \cite{8929906} for precisely describing the structure of most product-like codes with every variable node having degree two. Within this framework, the authors in \cite{8929906} also proposed tiled diagonal zipper codes which can be seen as a combination of continuously interleaved BCH codes \cite{Coe2012} and staircase codes \cite{Smith12}. In addition to spatial coupling, another line of work is to construct symmetry-based product codes \cite{7309002} to reduce the block size of product codes \cite{1057464} while having the same component code and similar code rates. With this property, one can employ stronger algebraic component codes to construct symmetry-based product codes in a bid to achieve better waterfall and error floor performance while maintaining similar block sizes and code rates as the conventional product codes. The first examples of such codes are half-product codes \cite{5671556}, whose codewords are derived from product codes with the additional constraint that the code arrays are anti-symmetric. Since each off-diagonal symbol of a half-product code array is repeated twice, the repeated symbols are punctured before transmission. Therefore, half-product codes have an effective blocklength about half to that of the product codes from which they are derived. Later, this idea inspired the design of quarter-product codes and octal-product codes in \cite{7282455} as well as half-braided BCH codes in \cite{7537380}. However, all the above symmetric-based product codes require square code blocks and the same component codes for row and column encoding. In addition, the codes in \cite{7282455} restrict the component codes to be reversible (i.e., a code that is invariant under a reversal of the coordinates in each codeword \cite{reversible1}). These restrictions reduce the design space of symmetric-based product codes and may limit their potential applications.

This paper focuses on designing new FEC schemes under low-complexity iBDD to achieve better waterfall and error floor performance with lower miscorrection probability than staircase codes \cite{Smith12}. Motivated by spatial coupling and symmetry, we propose sub-block rearranged staircase (SR-staircase) codes. The proposed codes can be constructed to have each code block with a size of $1/q$ to that of the conventional staircase codes with the same algebraic component codes while maintaining the same code rate, for any positive integer $q$. This means that we can employer strong algebraic component codes to construct SR-staircase codes with a similar or the same code block size and rate as staircase codes with weak component codes. The proposed SR-staircase codes have a flexible structure and offer larger degrees of freedom in code design compared to the conventional staircase codes and symmetric-based product codes. However, unlike all the aforementioned symmetric-based product codes, the proposed codes do not impose any additional constraint on the component codes and code array shapes. The main contributions of this paper are as follows.
\begin{itemize}
\item We propose SR-staircase codes which inherit the benefits from both symmetry and spatial coupling. We first introduce the code structure, encoding and decoding procedures. We then extend the proposed construction to a large coupling width. The connections between SR-staircase codes, conventional staircase codes and other spatially coupled codes are discussed.

\item We investigate the performance of the proposed codes under miscorrection-free iBDD (i.e., the component BDD only outputs either the correct codeword or the original received vector) on the BSC. By looking into the graph model, we first apply density evolution (DE) \cite{7890389} on SR-staircase codes with deterministic structures and characterize the decoding thresholds. We also derive a necessary condition under which the proposed codes achieve a larger decoding threshold than staircase codes. In addition, we investigate the error floor performance by analyzing the contributing error patterns and their multiplicities. Our results demonstrate that the decoding threshold and error floor of SR-staircase codes can be improved by using a large coupling width.

\item Numerical results are provided and show that the designed SR-staircase codes achieve better waterfall and error floor performance over staircase codes under iBDD. It is also interesting to note that the performance of the proposed codes under iBDD is very close to that under miscorrection-free iBDD due to the use of strong BCH component codes. We stress that the use of BCH component codes with stronger error correction capability can offer better error correction and error detection than employing BCH component codes with weaker error correction capability and extended parity bits, e.g., \cite{Smith12}.
\end{itemize}

\subsection{Notation}\label{sec:note}
This paper uses the following notations. Scalars, vectors and matrices are written in lightface, boldface and boldface capital letters, respectively, e.g., $x$, $\boldsymbol{x}$ and $\boldsymbol{X}$. The $r$-th row of a matrix $\boldsymbol{X}$ is represented by $\boldsymbol{x}_r \in \boldsymbol{X}$. $\mathbb{N}\triangleq \{1,2,\ldots\}$ represents the set of natural numbers. We define $[n] \triangleq \{1,\ldots,n\}$ for any $n\in \mathbb{N}$. $\left\lceil x\right\rceil$ gives the nearest integer that is not smaller than $x$ while $\lfloor x\rfloor $ gives the nearest integer that is not larger than $x$. We define a function $\varphi:\mathbb{N}  \rightarrow \{1,2\}$ as $\varphi(x) = \frac{3-(-1)^x}{2}$. The binary field and the collection of binary matrices of size $m\times n$ are denoted by $\mathbb{F}_2$ and $\mathbb{F}_2^{m,n}$, respectively. An $m \times n$ all-zero matrix is represented by $\boldsymbol{0}_{m,n}$. The transpose operation is denoted by $(.)^\mathsf{T}$. LCM denotes the least common multiple. The Hamming weight function is denoted by $w_{\mathsf{H}}(\cdot)$. For a set $\mathcal{S}$, $|\mathcal{S}|$ outputs its cardinality. For a length-$n$ vector $\boldsymbol{x}$, $\boldsymbol{x}(\mathcal{S})$ is a sub-vector of $\boldsymbol{x}$ by taking the elements in the positions of $\mathcal{S} \subseteq [n]$. The indicator function is represented by $\mathbbm{1}\{\cdot\}$.

\section{Sub-block Rearranged Staircase Codes}
In this section, we introduce the encoding and decoding of SR-staircase codes. We also discuss the relationship between the proposed codes and the conventional staircase codes \cite{Smith12}. In this work, we consider the underlying component codes to be binary primitive BCH codes. However, like the conventional staircase codes, the choice of the component codes for SR-staircase codes does not preclude other linear codes such as polar codes \cite{9083091} and LDPC codes \cite{9120519}.

\subsection{Encoding}
A SR-staircase code comprises a sequence of code blocks $\boldsymbol{B}_1,\boldsymbol{B}_2\ldots.$ At time $i\in \mathbb{N}$, code block $\boldsymbol{B}_i = [\boldsymbol{K}_i,\boldsymbol{P}_i]$ is a concatenation of information block $\boldsymbol{K}_i$ and parity block $\boldsymbol{P}_i$. To construct the SR-staircase code, two shortened BCH codes $\mathcal{C}_j$ for $j\in\{1,2\}$ are used. We denote by $k_j$, $n_j$, $t_j$, $e_j$, and $\boldsymbol{G}_j$ the message length, codeword length, error correction capability, shortening parameter, and generator matrix, respectively, of $\mathcal{C}_j$. Note that we can also express the codeword length and information length of $\mathcal{C}_j$ as $n_j = 2^{\nu_j}-1-e_j$ and $k_j = 2^{\nu_j}-1-\nu_j t_j-e_j$, respectively, for some positive integer $\nu_j \geq 3$, where $\nu_j$ is Galois field extension \cite[Ch. 3.3]{Lin09}.

The encoding of SR-staircase codes is performed in a recursive manner like the conventional staircase codes. The main difference is that each preceding SR-staircase code block $\boldsymbol{B}_{i-1}$ is required to be decomposed into $q_1$ equal-size sub-blocks if $i\in 2\mathbb{N}-1$ and $q_2$ equal-size sub-blocks if $i\in 2\mathbb{N}$. Each sub-block is then transposed before performing the component code encoding. The size of $\boldsymbol{B}_{i}$ is $\frac{m_1}{q_1}\times m_2$ if $i\in 2\mathbb{N}-1$ and $\frac{m_2}{q_2}\times m_1$ if $i\in 2\mathbb{N}$. Moreover, all the bits in each row of $\boldsymbol{B}_{i}$ are the last $m_2$ bits of a codeword of $\mathcal{C}_2$ when $i\in 2\mathbb{N}-1$ and the last $m_1$ bits of $\mathcal{C}_1$ when $i\in 2\mathbb{N}$. Note that the numbers of columns of $\boldsymbol{B}_i$, $m_1$ and $m_2$, have to be divisible by $q_1$ and $q_2$, respectively. We also denote by $w$ the coupling width, where $w \geq 2$ (i.e., $w=1$ means uncoupled) and both $m_1$ and $m_2$ have to be divisible by $w-1$. In the following, we present the encoding procedures.

\subsubsection{Case $w=2$}\label{subsec:enc}
For ease of presentation, we first describe the encoding steps for $i\in 2\mathbb{N}$.

\emph{\textbf{Step 1 (Initialization):}} Set all the entries of $\boldsymbol{B}_0$ to zero: $\boldsymbol{B}_0 = \boldsymbol{0}_{\frac{m_2}{q_2},m_1}$. $\boldsymbol{B}_0$ is known by the encoder and decoder pair. The recursive encoding process starts from $i=1$.

\emph{\textbf{Step 2 (Decomposition):}} The preceding block $\boldsymbol{B}_{i-1}$ with size $\frac{m_1}{q_1}\times m_2$ is divided into $q_2$ consecutive equal-size sub-blocks $\boldsymbol{B}_{i-1,1},\boldsymbol{B}_{i-1,2},\ldots,\boldsymbol{B}_{i-1,q_2}$. That is,
\begin{align}\label{eq:interleaver1}
\boldsymbol{B}_{i-1} = \left[\boldsymbol{B}_{i-1,1},\boldsymbol{B}_{i-1,2},\ldots,\boldsymbol{B}_{i-1,q_2}\right].
\end{align}
Each sub-block of $\boldsymbol{B}_{i-1}$ has size $\frac{m_1}{q_1}\times \frac{m_2}{q_2}$.

\emph{\textbf{Step 3 (Transformation):}}  Apply matrix transpose to each sub-block of $\boldsymbol{B}_{i-1}$ in Step 2 and combine them to form block $\boldsymbol{B}^{\pi}_{i-1}$ with size $\frac{m_2}{q_2} \times \frac{m_1 q_2}{q_1}$, given by
\begin{align}\label{eq:int1}
\boldsymbol{B}^{\pi}_{i-1}=\left[\boldsymbol{B}^\mathsf{T}_{i-1,1},\boldsymbol{B}^\mathsf{T}_{i-1,2},\ldots,\boldsymbol{B}^\mathsf{T}_{i-1,q_2}\right].
\end{align}
Each sub-block of $\boldsymbol{B}^{\pi}_{i-1}$ is of size $\frac{m_2}{q_2} \times \frac{m_1}{q_1}$. Note that all bits in the same column position of every transposed sub-block, $\boldsymbol{B}^\mathsf{T}_{i-1,1},\ldots,\boldsymbol{B}^\mathsf{T}_{i-1,q_2}$, belong to the same component codeword of $\mathcal{C}_2$. The transformation of $\boldsymbol{B}_{i-1}$ into $\boldsymbol{B}^{\pi}_{i-1}$ in \eqref{eq:int1} can be generalized by employing a permutation function $\pi(.)$ which permutes the rows and columns of a matrix, such that
\begin{align}\label{enc_random_int}
\boldsymbol{B}^{\pi}_{i-1}=\pi\left(\left[\boldsymbol{B}^\mathsf{T}_{i-1,1},\boldsymbol{B}^\mathsf{T}_{i-1,2},\ldots,\boldsymbol{B}^\mathsf{T}_{i-1,q_2}\right]\right).
\end{align}

\emph{\textbf{Step 4 (Array Concatenation):}} Arrange the information bits to be encoded for the $i$-th code block as an $\frac{m_2}{q_2} \times (k_1-\frac{m_1q_2}{q_1})$ block $\boldsymbol{K}_i$. Concatenate an all-zero block $\boldsymbol{0}_{\frac{m_2}{q_2},e_1}$ (representing shortened bits), the rearranged preceding block $\boldsymbol{B}^{\pi}_{i-1}$ from Step 3, and information block $\boldsymbol{K}_i$ to construct an $\frac{m_2}{q_2}\times (k_1+e_1)$ message matrix to be encoded at time $i$
\begin{align}\label{eq:K_prime1}
\boldsymbol{K}'_i = \left[\boldsymbol{0}_{\frac{m_2}{q_2},e_1},\boldsymbol{B}^{\pi}_{i-1},\boldsymbol{K}_i\right].
\end{align}

\emph{\textbf{Step 5 (Component Code Encoding):}} Perform row-by-row systematic component code encoding to obtain the codeword matrix with size $\frac{m_2}{q_2} \times n_1$ at time $i$
\begin{align}\label{eq:C1}
\boldsymbol{C}_i =& \boldsymbol{K}'_i\boldsymbol{G}_1  \nonumber \\
=&\left[\boldsymbol{0}_{\frac{m_2}{q_2},e_1},\boldsymbol{B}^{\pi}_{i-1},\boldsymbol{K}_i,\boldsymbol{P}_i\right] \nonumber \\
=&\left[\boldsymbol{0}_{\frac{m_2}{q_2},e_1},\boldsymbol{B}^{\pi}_{i-1},\boldsymbol{B}_i\right],
\end{align}
where $\boldsymbol{P}_i$ is the parity block with size $\frac{m_2}{q_2} \times (n_1-k_1)$. Finally, $\boldsymbol{B}_{i}=[\boldsymbol{K}_i,\boldsymbol{P}_i]$ is an $\frac{m_2}{q_2}\times m_1$ code block that will be transmitted. Each row of $[\boldsymbol{B}^{\pi}_{i-1},\boldsymbol{B}_i]$ is a shortened codeword of $\mathcal{C}_1$.

The encoding steps to obtain $\boldsymbol{B}_i$ for $i\in 2\mathbb{N}-1$ are similar to the above. After Step 5, each row of $[\boldsymbol{B}^{\pi}_{i-1},\boldsymbol{B}_i]$ is a shortened codeword of $\mathcal{C}_2$ for $i\in 2\mathbb{N}-1$. The relation between the component codeword length, time index $i$, the number of decomposed sub-blocks in $\boldsymbol{B}_i$ and the number of columns of $\boldsymbol{B}_i$ satisfies
\begin{align}\label{eq:con1}
n_{\varphi(i)} = m_{\varphi(i)}+\frac{m_{\varphi(i)}\cdot q_{\varphi(i-1)}}{q_{\varphi(i)}},
\end{align}
where $\varphi(x) = \frac{3-(-1)^x}{2}$ as defined in Sec. \ref{sec:note}. The code rate is given by
\begin{align}\label{eq:rate}
R =& \frac{1}{2}\left(\frac{k_1}{m_1}+\frac{k_2}{m_2}-\frac{q_2}{q_1}-\frac{q_1}{q_2}\right) \nonumber \\
=&1-\frac{1}{2}\left(\frac{\nu_1t_1}{m_1}+\frac{\nu_2t_2}{m_2}\right).
\end{align}

Alternatively, SR-staircase codes can be described by using the zipper code framework \cite{8929906}, where $\boldsymbol{B}^{\pi}_{i-1}$ is the virtual buffer and $\boldsymbol{B}_{i-1}$ is the corresponding real buffer. The transformation of $\boldsymbol{B}_{i-1}$ into $\boldsymbol{B}^{\pi}_{i-1}$ in Step 3 can be described by using a bijective mapping function.

\begin{figure}[t!]
	\centering
\includegraphics[width=2.4in,clip,keepaspectratio]{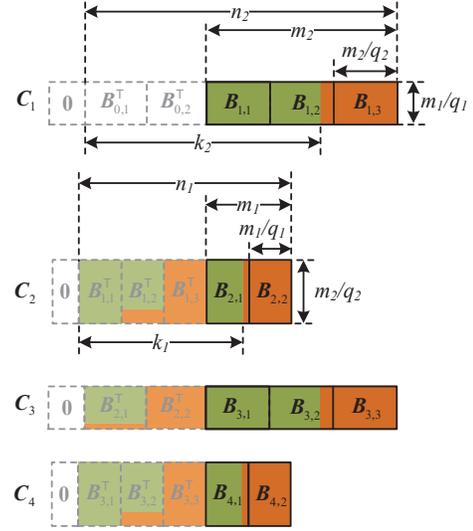}
\caption{Illustration of a SR-staircase code with $w=2$ and $(q_1,q_2)=(2,3)$ whose codewords are in the form of row codewords only. Sub-blocks contain zeros (white), information bits (green), and parity bits (brown) are shown}
\label{fig:s_staircase_2a}
\end{figure}

\begin{example}\label{example1}
Consider a SR-staircase code with $w=2$, $q_1 = 2$ and $q_2 = 3$. The codeword matrices obtained in \eqref{eq:C1} from  Step 5 are shown in Fig. \ref{fig:s_staircase_2a}. The sub-blocks indicated by light colors and gray dash lines are involved in the encoding but will not be transmitted. Specifically, Fig. \ref{fig:s_staircase_2a} shows the codeword matrix $\boldsymbol{C}_i$ in the form of row codewords only. Each row in $[\boldsymbol{B}^{\pi}_{i-1},\boldsymbol{B}_i]$ is a shortened codeword of $\mathcal{C}_1$ and $\mathcal{C}_2 $ for $i$ being even and odd, respectively.
\demo
\end{example}

\subsubsection{Case $w>2$}\label{subsec:enc_w3}
In this case, we need to ensure that each sub-block used for coupling has the same size. This is possible if and only if $m_1=m_2 \triangleq m$ and $q_1 = q_2 \triangleq q$. Consider $i\in2\mathbb{N}$. To obtain $\boldsymbol{B}_i$, we first modify Step 1 of the encoding in Sec. \ref{subsec:enc} by setting $\boldsymbol{B}_0,\ldots,\boldsymbol{B}_{w-2}$ to all-zero matrices. Next, we modify Step 4 by further dividing the transformed preceding code block $\boldsymbol{B}^{\pi}_{i-l}$ obtained from \eqref{eq:int1} into $w-1$ consecutive equal-size sub-blocks for $l\in[w-1]$
\begin{align}\label{enc:large_w}
\boldsymbol{B}^{\pi}_{i-l} = \left[\boldsymbol{B}^{\pi}_{i-l,1},\ldots,\boldsymbol{B}^{\pi}_{i-l,w-1}\right],
\end{align}
where each sub-block is an $\frac{m}{q}\times \frac{m}{w-1}$ binary matrix. For $l\in[w-1]$, the $l$-th sub-block of preceding transformed code block $\boldsymbol{B}^{\pi}_{i-l}$, i.e., $\boldsymbol{B}^{\pi}_{i-l,l}$, is used for constructing the message matrix at time $i$
\begin{align}\label{enc:large_w1}
\boldsymbol{K}'_i = \left[\boldsymbol{0}_{\frac{m}{q},e_1},\boldsymbol{B}^{\pi}_{i-1,1},\boldsymbol{B}^{\pi}_{i-2,2},\ldots,\boldsymbol{B}^{\pi}_{i-w+1,w-1},\boldsymbol{K}_i\right].
\end{align}
Thus, \eqref{eq:K_prime1} in Step 4 is replaced by \eqref{enc:large_w1}. The rest of the encoding steps are the same as those in Sec. \ref{subsec:enc}. The overall code rate does not change with $w$.

It is important to note that when $w\geq q+1$, the bits in different column positions of the coupled block $[\boldsymbol{B}^{\pi}_{i-1,1},\ldots,\boldsymbol{B}^{\pi}_{i-w+1,w-1}]$ are protected by different component codewords because any pair of sub-blocks, $\boldsymbol{B}^{\pi}_{i-l,l}$ and $\boldsymbol{B}^{\pi}_{i-l',l'}$ with $l\neq l'$ and $l,l' \in [w-1]$, are decomposed from different preceding code blocks.

\begin{remark}
Although we only consider using the same component code across the rows of each SR-staircase code block in this work for simplicity, it is possible to use component code mixtures such that the component code varies among the rows of the same code block. In fact, this was suggested for the conventional staircase codes in \cite[Sec. 4.4]{BPSmith11}. However, it was proved that employing component code mixtures is not beneficial to the asymptotic performance of spatially coupled product codes \cite{7541672}. In addition, one may also use the proposed technique to construct uncoupled product codes in order to employ strong BCH component codes. In this case, the `checks on checks' array on the resultant product codes will become different depending on whether rows or columns are encoded first. As a result, the parity bits of the `checks on checks' array can only be protected by either column or row codewords, leading to some loss in performance.
\end{remark}

\subsection{Connections to Other Spatially Coupled Codes}
\begin{figure}[t!]
	\centering
\includegraphics[width=3.2in,clip,keepaspectratio]{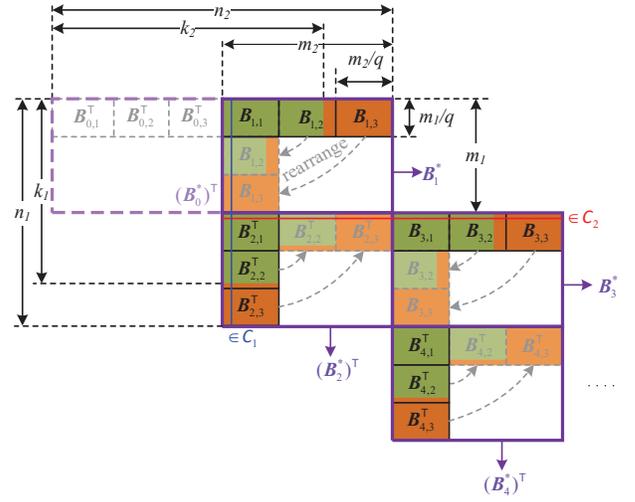}
\caption{The relation between a SR-staircase code with $q=3$ and a staircase code (purple). The rearrangement of the sub-blocks of the SR-staircase code and two component codewords that belong to $\mathcal{C}_1$ (blue) and $\mathcal{C}_2$ (red), respectively, are indicated.}
\label{fig:compare1}
\end{figure}

SR-staircase codes are motivated and derived by introducing symmetry in the conventional staircase codes  \cite{Smith12}. Consider the SR staircase code in Sec. \ref{subsec:enc} with $w=2$ and let $q \triangleq q_1 = q_2$. By concatenating $q$ identical SR-staircase code block $\boldsymbol{B}_i$, one obtains the resultant staircase code block at time $i$ as
\begin{align}\label{eq:B1}
\boldsymbol{B}^*_i =& \left[\boldsymbol{B}^\mathsf{T}_i,\ldots,\boldsymbol{B}^\mathsf{T}_i\right]^\mathsf{T} \nonumber\\
=& \left[[\boldsymbol{K}_i,\boldsymbol{P}_i]^\mathsf{T},\ldots,[\boldsymbol{K}_i,\boldsymbol{P}_i]^\mathsf{T}\right]^\mathsf{T} \nonumber\\
= & \left[\left[\boldsymbol{K}^\mathsf{T}_i,\ldots,\boldsymbol{K}^\mathsf{T}_i\right]^\mathsf{T},\left[\boldsymbol{P}^\mathsf{T}_i,\ldots,\boldsymbol{P}^\mathsf{T}_i\right]^\mathsf{T}\right] \nonumber\\
=&[\boldsymbol{K}^*_i,\boldsymbol{P}^*_i].
\end{align}
where $\boldsymbol{K}^*_i  = [\boldsymbol{K}^\mathsf{T}_i,\ldots,$ $\boldsymbol{K}^\mathsf{T}_i]^\mathsf{T}$ and $\boldsymbol{P}^*_i  = [\boldsymbol{P}^\mathsf{T}_i,\ldots,\boldsymbol{P}^\mathsf{T}_i]^\mathsf{T}$ are the $i$-th information block and parity block of staircase codes \cite{Smith12}. Consider $i\in 2\mathbb{N}$. The sizes of the current and preceding staircase code blocks satisfy $\boldsymbol{B}^*_i \in \mathbb{F}_2^{m_2,m_1}$ and $\boldsymbol{B}^*_{i-1} \in \mathbb{F}_2^{m_1,m_2}$. Notice that $\boldsymbol{B}^*_{i-1}$ can be rearranged into $\boldsymbol{B}^{\pi*}_{i-1} \in \mathbb{F}_2^{m_2,m_1}$, which consists of $q$ identical rearranged code blocks $\boldsymbol{B}^{\pi}_{i-1}\in \mathbb{F}_2^{\frac{m_2}{q},m_1}$, i.e.,
\begin{align}\label{eq:B2}
\boldsymbol{B}^{\pi*}_{i-1} = \left[\left(\boldsymbol{B}^{\pi}_{i-1}\right)^\mathsf{T},\ldots,\left(\boldsymbol{B}^{\pi}_{i-1}\right)^\mathsf{T}\right]^\mathsf{T},
\end{align}
where the construction of $\boldsymbol{B}^{\pi}_{i-1}$ follows from either \eqref{enc_random_int} or \eqref{eq:int1}. As a result, each row of $[\boldsymbol{B}^{\pi*}_{i-1},\boldsymbol{B}_{i}]$ is a valid codeword of $\mathcal{C}_1$. Clearly, it can be seen that each code block $\boldsymbol{B}^*_i$ is drawn from a subset of the set of the code blocks of staircase codes due to symmetry, i.e., having $q-1$ replicas of $\boldsymbol{B}_i$. Thus, the resultant staircase code $\boldsymbol{B}^*_1,\ldots$ is a subcode of the conventional staircase code. Notice that when $q=1$, the encoding steps in Sec. \ref{subsec:enc} produce the conventional staircase codes. By removing any $q-1$ replicas of $\boldsymbol{B}_i$ as they do not contain any new information, the resultant SR-staircase codes achieve the same rates and an effective code block size of $1/q$ to the conventional staircase codes from which they are derived. In this regard, the proposed construction enables to employ stronger BCH codes to construct SR-staircase codes with improved error performance while maintaining a similar or the same code block size and rate compared to staircase codes. To visualize the relationships in \eqref{eq:B1} and \eqref{eq:B2}, we show the code blocks of a SR-staircase code with $q=3$ and a staircase code in Fig. \ref{fig:compare1}, where both codes use the same component codes and have the same rate. Note that the colors follow a similar style as in Fig. \ref{fig:s_staircase_2a}. Clearly, the SR-staircase code has a block size of $1/3$ to that of the benchmark staircase code.

The proposed SR-staircase codes are also close to tiled diagonal zipper codes \cite[Sec. IV-E]{8929906}. Specifically, tiled diagonal zipper codes can be seen as a special case of the proposed SR-staircase codes by fixing $\mathcal{C}_1=\mathcal{C}_2$, $w-1=q_1=q_2$, $m_1=m_2$, and a specific permutation of \eqref{enc_random_int}. However, we emphasize that the proposed SR-staircase codes are motivated and derived by applying the idea of symmetry from symmetric-based product codes \cite{7309002,7282455} to staircase codes \cite{Smith12} as illustrated above. Compared to tiled diagonal zipper codes, the proposed codes have more code parameters, such as the decomposition number $(q_1, q_2)$ and coupling width $w$, which are explicitly defined and play very important roles in determining the rate, code block size, and performance. This, together with the capability of using a pair of different component codes $(\mathcal{C}_1,\mathcal{C}_2)$, give rise to more flexible code structures for SR-staircase codes. Thus, the proposed codes can be constructed to meet a wider range of requirements. We emphasize that the aim of this work is to design codes with superior waterfall and error floor performance over staircase codes under iBDD. To this end, we use rigorous density evolution and error floor analysis to design code parameters $(w,q_1,q_2,m_1,m_2)$ and justify the choice of component codes $(\mathcal{C}_1,\mathcal{C}_2)$.

The proposed SR-staircase codes are also related to the class of partially coupled codes, i.e., \cite{PIC2020,GSCPCC2021ISIT,9491085}, recently proposed by us in the sense that a fraction of information and/or parity bits in one code block are coupled and become a part of the input to the encoders of consecutive code blocks. These bits are repeated before coupling and component code encoding while all repeated bits are punctured before transmission. This allows us to introduce stronger component codes to improve the overall decoding performance of coupled codes.

\subsection{Decoding}\label{sec:dec}
The decoding of SR-staircase codes is performed in a sliding window fashion, similar to staircase codes. To avoid repetition, we only point out the main difference. We denote by $W$ the decoding window size satisfying $W>w$ and $\boldsymbol{Y}_{i}$ the received code block corresponding to $\boldsymbol{B}_i$ after hard-decision demapping. Consider $m_1 = m_2 \triangleq m$ and $q_1 = q_2 \triangleq q$ for simplicity. The decoder constructs the received codeword matrix corresponding to $\boldsymbol{C}_i$ in \eqref{eq:C1}
\begin{align}\label{eq:Dl-1w}
\boldsymbol{D}_{i} = \left[\boldsymbol{0}_{\frac{m}{q},e_{\varphi(i)}},\boldsymbol{Y}^{\pi}_{i-1,1},\boldsymbol{Y}^{\pi}_{i-2,2},\ldots,\boldsymbol{Y}^{\pi}_{i-w+1,w-1},\boldsymbol{Y}_{i}\right],
\end{align}
where $\boldsymbol{Y}^{\pi}_{i-l,l},l\in [w-1]$ is the $l$-th sub-block decomposed from $\boldsymbol{Y}^{\pi}_{i-l}$, which is obtained by applying the transformation of \eqref{eq:int1} to $\boldsymbol{Y}_{i-l}$. Then, BDD is applied to each row of $\boldsymbol{D}_{i}$ with non-zero syndrome and the rest of the decoding steps directly follow those in \cite[Sec. IV-A]{Smith12}.

In this work, we restrict the decoding to be iBDD due to its simplicity and low complexity. In Section \ref{sec:sim}, we will show that iBDD is suffice for SR-staircase codes to operate close to miscorrection-free performance as a result of using component codes with large $(t_1,t_2)$. We note that a range of decoding algorithms, e.g., \cite{6831429,hager2017approaching,8463624,8856224,8832202} have been proposed for product-like codes to bring their decoding performance close to miscorrection-free performance or beyond at the cost of increased complexity. Hence, it is also beneficial to apply these decoding algorithms to SR-staircases. This will be investigated in our future work.

\section{Decoding Threshold Analysis}\label{sec:pa}
In this section, we analyze the decoding thresholds of SR-staircase codes by using DE. Based on the analysis, we then present a guideline for designing the parameters for SR-staircase codes to achieve a better threshold than the conventional staircase codes.

\begin{figure*}[t!]
	\centering
\includegraphics[width=6.5in,clip,keepaspectratio]{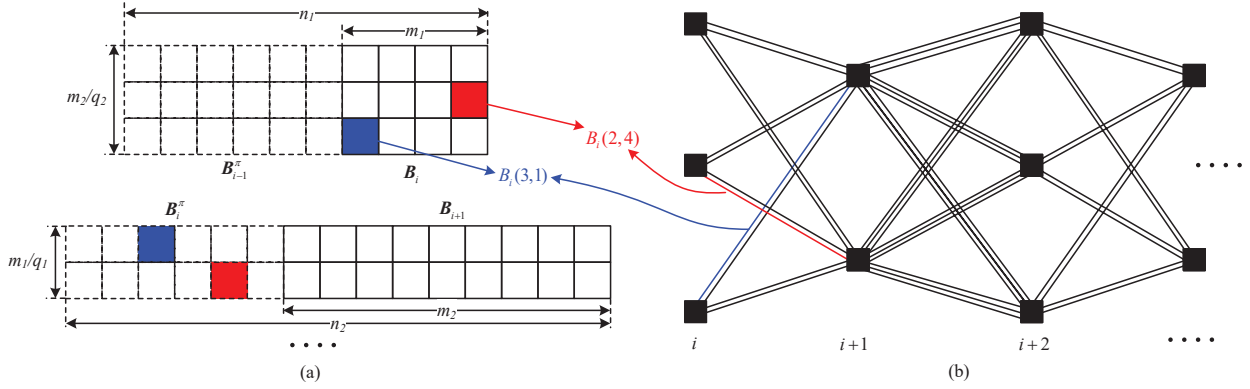}
\caption{Illustrations of a SR-staircase code with $w=2$, $(m_1,m_2) = (4,9)$ and $(q_1,q_2) = (2,3)$ with $i \in 2\mathbb{N}$. (a) SR-staircase code blocks (dash lines and solid lines illustrate preceding and current code blocks, respectively); (b) Tanner graph representation.}
\label{fig:graph_w2}
\end{figure*}

\subsection{Graph Model}\label{sec:graph_model}
We first study the graph model of the proposed codes. Following the approach in \cite{7890389}, we consider a deterministic code structure since the interleaver of the proposed codes is fixed. The analysis performed on a deterministic code structure allows one to make precise statements about the performance of actual codes. Although one can employ random interleaving in the proposed codes as shown in \eqref{enc_random_int}, the deterministic code structures often give rise to implementation advantages over random ensembles.

For ease of understanding, we first consider the case of $w=2$. From Sec. \ref{subsec:enc}, we know that code block $\boldsymbol{B}_i$ has $\frac{m_2}{q_2}$ rows for $i\in 2\mathbb{N}$ and $\frac{m_1}{q_1}$ rows for $i\in 2\mathbb{N}-1$. By using the Tanner graph representation \cite{1056404}, it can be seen that the $i$-th spatial position (time instance) on the graph has $\frac{m_2}{q_2}$ check nodes (CNs) when $i\in 2\mathbb{N}$ and $\frac{m_1}{q_1}$ CNs when $i\in 2\mathbb{N}-1$ because one component codeword poses constraints on a row of $\boldsymbol{B}_i$. Each bit in $\boldsymbol{B}_i$ is represented by a variable node (VN) that connects a pair of CNs in the $i$-th and $(i+1)$-th spatial positions via an edge. Thus, each VN always has degree 2. All CNs in any two neighboring spatial positions are fully connected. More precisely, each pair of CNs in the two neighboring spatial positions $i$ and $i+1$, are connected via $q_1$ and $q_2$ edges for $i\in2\mathbb{N}$ and $i\in2\mathbb{N}-1$, respectively, where a VN lies on each edge. We use an example to illustrate the graph representation of a SR-staircase code with given specific parameters.

\begin{example}\label{exam1}
Consider a SR-staircase code with $(m_1,m_2) = (4,9)$ and $(q_1,q_2) = (2,3)$. The code blocks and the corresponding graph model of this SR-staircase code are shown in Fig. \ref{fig:graph_w2}(a) and Fig. \ref{fig:graph_w2}(b), respectively. Consider $i \in 2\mathbb{N}$. Since each VN always has degree 2, we use an edge to represent a VN that connects a pair of CNs for simplicity. We label two bits in $\boldsymbol{B}_i$, i.e., $B_i(3,1)$ and $B_i(2,4)$, in Fig. \ref{fig:graph_w2}(a) and mark their corresponding edges (VNs) in the Tanner graph with the same color in Fig. \ref{fig:graph_w2}(b). Note that the code structure and graph model in Fig. \ref{fig:graph_w2} are based on the transformation in \eqref{eq:int1}. If a random permutation function in \eqref{enc_random_int} is adopted, the bit label of each edge in Fig. \ref{fig:graph_w2}(b) will change while the connectivity between CNs remains unchanged.
\demo
\end{example}

We now consider the case of $w>2$ and using the coupling pattern shown in \eqref{enc:large_w} and \eqref{enc:large_w1}. We set $m_1 = m_2 \triangleq  m$ and $q_2=q_2 \triangleq  q$ by following Sec. \ref{subsec:enc_w3}. Different from $w=2$, all CNs in spatially positions $i$ and $i+l,\forall l\in[w-1]$, are fully connected. When $w\geq q+1$, each pair of CNs in two coupling spatial positions, $i$ and $i+l$, are connected by only one edge. This is because the bits in different column positions of $[\boldsymbol{B}^{\pi}_{i-1,1},\ldots,\boldsymbol{B}^{\pi}_{i-w+1,w-1}]$ are protected by different component codewords according to Sec. \ref{subsec:enc_w3}. As a result, each bit in $\boldsymbol{B}_i$ is protected by two component codewords. It should be noted that this may not hold in general if the coupling pattern is completely random. In contrast, when $w=2$ (and we still assume $m_1 = m_2 \triangleq m$ and $q_2=q_2 \triangleq q$), the bits in the same column position of every coupled sub-block $\boldsymbol{B}^{\pi}_{i-1,1},\ldots,\boldsymbol{B}^{\pi}_{i-1,q}$ are protected by the same component codeword. Hence, the SR-staircase code with $w=2$ has a multi-edge graph representation shown in Fig. \ref{fig:graph_w2}(b) such that every $q$ bits are protected by two component codewords. When $2<w<q+1$, the connectivity between CNs is mixed with single-edge and multi-edge. For this case, the number of connecting edges ranges from 1 to $\lceil\frac{q}{w-1}\rceil$ and depends specifically on the values of $q$ and $w$.

\subsection{Density Evolution}
We derive the DE equations for the BSC based on the graph model in \eqref{sec:graph_model}. We note that various techniques were introduced in the literature to analyze the performance of product-like codes \cite{7954697,7890389,8082562}. Both \cite{7954697} and \cite{8082562} applied DE to the ensembles that can represent a range of product-like codes. To make precise statements about the performance of the proposed codes with deterministic structures under iBDD, we adopt the approach in \cite{7890389} to perform DE analysis. Moreover, we assume that the underlying BDD is miscorrection-free as it is a necessary condition to conduct the DE analysis \cite{7890389,8082562}.

\subsubsection{}
We start with the case of $w=2$. Consider the SR-staircase code constructed in Sec. \ref{subsec:enc} with code blocks $\boldsymbol{B}_i,i\in[L]$. Let $p$ be the crossover probability of a BSC. We define the effect channel quality to be
\begin{align}\label{eq:MP}
M_{\varphi(i)} \triangleq p n_{\varphi(i)} \overset{\eqref{eq:con1}}{=}p \left(m_{\varphi(i)}+\frac{m_{\varphi(i)}\cdot q_{\varphi(i-1)}}{q_{\varphi(i)}}\right),
\end{align}
whose operational meaning is the expected average number of bits received in errors per component code constraint of $\mathcal{C}_{\varphi(i)}$ and $\varphi(.)$ is a mapping function defined in Sec. \ref{sec:note}. Hence, we are interested in the probability that a CN declares a decoding failure after $\ell$ iterations as $n_{\varphi(i)} \rightarrow \infty$. To track this probability as a function of $\ell$, we define a parameter $x^{(\ell)}_i, i\in[L]$, whose operational meaning is that the probability of a randomly chosen erroneous bit attached to a component code of $\mathcal{C}_{\varphi(i)}$ in $\boldsymbol{B}_i$ is not recovered after $\ell$ decoding iterations converges asymptotically to $x^{(\ell)}_i$. The bit will not be recovered if its attached component codeword has more than $t_{\varphi(i)}$ errors. According to \cite{7890389}, the total number of errors in $\boldsymbol{B}_i$ per component code constraint at the start of the $\ell$-th iteration converges to a Poisson random variable with mean $\frac{M_{\varphi(i)}\left(x^{(\ell)}_{i-1}+x^{(\ell-1)}_{i+1}\right)}{2}$ as $n_{\varphi(i)} \rightarrow \infty$, where the error probabilities $x^{(\ell)}_{i-1}$ and $x^{(\ell-1)}_{i+1}$ are taken into account due to coupling. To characterize the iterative decoding process, one can first look at the error graph obtained from the corresponding Tanner graph, where all the VNs associated with the correctly received bits and their connected edges are removed. Then, the decoding of $\boldsymbol{B}_i$ is equivalent to removing any vertex in spatial position $i$ and its edges connected to the vertices in position $i+1$ if the number of those edges is no larger than $t_{\varphi(i)}$. As a result, the iterative decoding is characterized by a recursive complementary Poisson cumulative distribution function. Note that since $q_{\varphi(i)}$ is fixed and $n_{\varphi(i)}\gg q_{\varphi(i)}$, the above properties hold regardless of whether the Tanner graph is single-edge or multi-edge. For notation simplicity, we define $f(\lambda,t)\triangleq 1-\sum_{i=1}^{t-1}\frac{\lambda^i}{i!}e^{-\lambda}$ to be the complementary Poisson cumulative distribution function for a Poisson random variable $\lambda$ with support $t$. The DE equation for SR-staircase codes is
\begin{align}\label{eq:DEw2}
x^{(\ell)}_i =f\left(\frac{M_{\varphi(i)}}{2}\left(x^{(\ell)}_{i-1}+x^{(\ell-1)}_{i+1}\right),t_{\varphi(i)}\right),
\end{align}
where $x^{(0)}_i = 1$ for $i \in [L]$ and $x^{(\ell)}_i = 0$ for $i<1$ and $i>L$. The BSC decoding threshold is defined as $\bar{p} \triangleq \sup\left\{p>0\left|\lim_{\ell \rightarrow \infty}\boldsymbol{x}^{(\ell)} = \boldsymbol{0}_L \right.\right\}$.

\subsubsection{}
When $w > 2$, we have $m_1=m_2\triangleq m$ and $q_1=q_2 \triangleq q$ according to Sec. \ref{subsec:enc_w3}. In this case, the expected number of initial errors per component code is $M_1 = M_2 \triangleq M$. Recall that the $l$-th sub-block of preceding code block $\boldsymbol{B}^{\pi}_{i-l}$ for $l\in[w-1]$ is used as a part of the inputs to encode $\boldsymbol{B}_{i}$. Similarly, $\boldsymbol{B}_{i}$ is also used as a part of the inputs to encode $\boldsymbol{B}_{i+1},\ldots,\boldsymbol{B}_{i+w-1}$. The DE equation in \eqref{eq:DEw2} is then modified into
\begin{align}\label{eq:DEw3}
x^{(\ell)}_i =f\left(\frac{M}{2(w-1)}\sum\nolimits_{j=1}^{w-1}\left(x^{(\ell)}_{i-j}+x^{(\ell-1)}_{i+j}\right),t_{\varphi(i)}\right).
\end{align}

\subsubsection{Windowed Decoding}
The DE analysis above assumes that the decoding is performed for the entire spatial code chain. It is easy to extend the DE analysis to sliding window decoding. Consider a window size $W$ satisfying $w < W < L$. Then, the DE equation is modified into
\begin{align}\label{eq:DE_win}
x^{(\ell)}_i =\left\{\begin{array}{ll} \text{RHS of } \left\{\begin{array}{l}\eqref{eq:DEw2},\text{if } w=2\\
\eqref{eq:DEw3},\text{if } w>2
\end{array}\right., &i\in \{i',\ldots,i'+W-1\}\\
x^{(\ell-1)}_i,  &\text{otherwise}
\end{array}\right.,
\end{align}
where $i'\in [L-W+1]$ indicates the window position on the coupled code chain. It is important to note that under sliding window decoding, the error probability of coupled codes predicted by DE cannot reach 0 \cite{6374679}. In this case, the definition of BSC decoding threshold should be modified by accounting for a target error probability $\epsilon>0$ such that it becomes $\bar{p} \triangleq \sup\left\{p>0\left|\lim_{\ell \rightarrow \infty}x_i^{(\ell)} \leq \epsilon, \forall i\in[L] \right.\right\}$. However, to accurately compute the threshold for $w>2$, the window size needs to be very large for the decoding wave to form \cite{6374679,Farooq2021}.

\subsection{Decoding Threshold Results}\label{sec:DE_result}
In this section, we use the DE equations to characterize the decoding threshold of SR-staircase codes under full decoding of the entire spatial code chain. We first investigate the effective channel quality $\bar{M} \triangleq \sup\{M>0|\lim_{\ell \rightarrow \infty}\boldsymbol{x}^{(\ell)} = \boldsymbol{0}_L \}$ for SR-staircase codes with $m_1 = m_2\triangleq m$ and $q_1=q_2 \triangleq q$. This is because for given $(t_1,t_2,w)$, $\bar{M}$ becomes deterministic and will come in handy for quickly determining the BSC threshold of SR-staircase codes for various $(m,q)$. The results of $\bar{M}$ are reported in Table \ref{thres1}.

\begin{table*}[t!]
  \centering
 \caption{Decoding thresholds of SR-Staircase Codes in terms of effective channel quality $\bar{M}$}\label{thres1}
\begin{tabular}{c| c c c c c c c c c c c}
\hline
$(t_1,t_2)$ & $(2,2)$ & $(3,3)$ & $(4,4)$ & $(5,5)$ & (5,6)  & (6,6) & $(7,7)$ & $(7,8)$& $(8,8)$ & $(9,9)$ & $(10,10)$   \\
\hline
$w=2$ & 3.5880 & 5.7544 & 7.8397 & 9.8860 & 10.8607 & 11.9087 & 13.9148&14.8693 & 15.9082 & 17.8908 & 19.8641 \\
$w=3 $& 3.5880 & 5.7548 & 7.8428 & 9.8952 & 10.8762 & 11.9280 & 13.9488& 14.9007 & 15.9618 & 17.9692 & 19.9725\\
$w=4 $& 3.5880 & 5.7548 & 7.8429 & 9.8954 & 10.9006 & 11.9287 & 13.9507& 14.9434& 15.9654 & 17.9753 & 19.9821 \\
$w=5$ & 3.5880 & 5.7548 & 7.8429 & 9.8954 & 10.9028 & 11.9287 & 13.9507&14.9517 & 15.9655  & 17.9756 & 19.9827 \\
$w=6$ & 3.5880 & 5.7548 & 7.8429 & 9.8954 & 10.9040& 11.9287& 13.9507& 14.9542 & 15.9655 & 17.9756& 19.9827 \\
 \hline
\end{tabular}
\end{table*}

It can be seen that the effective channel quality improves with $t_1,t_2$ and $w$. Notice that $\bar{M} \leq t_1+t_2$, which is the necessary condition to guarantee successful decoding \cite{7890389}. As both $t_1$ and $t_2$ become large, $\bar{M}$ is getting closer to the $t_1+t_2$ upper bound when $w$ is large. Hence, it is more beneficial to use a large coupling width for a SR-staircase code with large $(t_1,t_2)$ than that with small $(t_1,t_2)$. Compared to the setting with $t_1 = t_2$, the one with $t_1 \neq t_2$ requires a larger $w$ for $\bar{M}$ to achieve its maximum value. Furthermore, it is interesting to note that this maximum value coincides with the potential threshold \cite{6887298} of the GLDPC ensemble with $t$ error correcting constituent BCH codes \cite[Table III]{7954697}. This implies that choosing a reasonable coupling width, e.g., $w=5$, is sufficient for the proposed codes to achieve the best possible threshold.

The BSC threshold $\bar{p}$ can then be easily determined by using $\bar{p} = \frac{\bar{M}}{2m}$ from \eqref{eq:MP}. The following theorem provides a necessary condition for SR-staircase codes to achieve a higher rate and BSC threshold, and smaller block size than staircase codes when $t_1 = t_2 \triangleq t$.

\begin{theorem}\label{prop1}
Consider a rate-$R'$ staircase code with given parameters $(m',\nu',t')$, decoding threshold $\bar{p}'$ and effective channel quality $\bar{M}'$. Consider a SR-staircase code with given parameters $w$, $q$, $\nu\geq \nu'$, $t>t'$, and the corresponding effective channel quality $\bar{M}$. Define $\beta\triangleq \text{LCM}(w-1,q)$ and let $R$, $\bar{p}$, and $m$ represent the SR-staircase code's rate, threshold and block size, respectively. If
\begin{align}\label{eq:req1}
a\triangleq\left\lceil\frac{t\nu m'}{t'\nu' \beta} \right\rceil  <  \min\left\{\sqrt{q}m',\frac{\bar{M}}{\bar{M}'}m',\frac{2^{\nu}-1}{2}\right\}\triangleq b,
\end{align}
then $\exists m\in[\beta a,b)\cap \beta\mathbb{Z}$ such that the resultant SR-staircase code has $R \geq R'$, $\bar{p}>\bar{p}'$ and $\frac{m^2}{q} \leq (m')^2$.
\end{theorem}
\begin{IEEEproof}
See Appendix \ref{app1}.
\end{IEEEproof}

Given a benchmark staircase code, using Theorem \ref{prop1}, we can quickly determine whether it is possible to construct a SR-staircase code with stronger BCH component codes to achieve $R \geq R'$, $\bar{p}>\bar{p}'$ and $\frac{m^2}{q} \leq (m')^2$. Once all the conditions in Theorem \ref{prop1} are fulfilled, we can simply choose $m = \beta\lceil\frac{tvm'}{t'\nu'\beta} \rceil$ based on \eqref{eq:req1}. This is because a smaller $m$ always gives rise to a larger BSC threshold for given $(t,w)$ due to the relation $\bar{p} = \frac{\bar{M}}{2m}$. Note that $\bar{M}$ is deterministic when given $(t,w)$ (see Table \ref{thres1}). Thus, the block size of the SR-staircase codes that achieves the aforementioned three goals can be determined by Theorem \ref{prop1} without searching. In addition, Theorem \ref{prop1} also implies that even employing stronger BCH component codes, it is still impossible to construct a staircase code to achieve a strictly larger BSC threshold without reducing its rate and increasing its block size. To see this, using the fact $w-1=q=1$ in \eqref{eq:req1}, we get the condition for such a staircase code to exist, which is $\lceil\frac{t\nu m'}{t'\nu'} \rceil <\min\{m',\frac{2^\nu -1}{2}\}$. Since $m'\leq \frac{2^{\nu'}-1}{2}\leq\frac{2^{\nu}-1}{2}$, we further obtain that $\lceil\frac{t\nu m'}{t'\nu'} \rceil  < m'$. However, this is contradictory to the conditions $t>t'$ and $\nu \geq \nu'$, which are introduced from employing stronger BCH component codes. In contrast, the proposed SR-staircase codes can achieve a strictly larger threshold without rate and block size penalties. It is possible to relax the rate requirement in Theorem \ref{prop1} by introducing a small variable $\delta \in (0,1)$ to allow SR-staircase codes to achieve a rate close to the benchmark staircase codes, i.e., $R\geq R'-\delta$. When either $t_1 \neq t_2$ or $q_1\neq q_2$, a search is required to find the optimal $m_1$ and $m_2$ that give the largest threshold.

We take several staircase codes in the literature as baselines and design SR-staircase codes with better thresholds, same or comparable rates, and smaller block sizes by using Theorem \ref{prop1}. The parameters of the designed codes and the corresponding benchmark staircase codes are reported in Table \ref{table2}. For illustrative purposes, we consider $\nu_1=\nu_2 \triangleq \nu $. Since only hard channel output is used, the BSC threshold can be equivalently converted into the additive white Gaussian noise (AWGN) threshold.

\begin{table*}[t!]
  \centering
   \begin{threeparttable}
 \caption{Decoding thresholds of SR-staircase Codes}\label{table2}
\begin{tabular}{c c c c c c c c c c}
\hline
Scheme & Rate  & $w$ & $\nu$ & $(m_1,m_2)$ & $(t_1,t_2)$   & $(q_1,q_2)$ & Block size & $\bar{p}$  & $\text{E}_\text{b}/\text{N}_0$ (dB)    \\\hline
\cite[Table I]{6787025} &0.9412 & 2& 11 & $(748,748)$ & $(4,4)$ & $(1,1)$ & 559504 & $5.2404\cdot 10^{-3}$ & 5.4163  \\\hdashline
 &0.9412 & 2& 11 & $(936,936)$ & $(5,5)$ & $(2,2)$ & 436178 & $5.2810\cdot 10^{-3}$ & 5.4069  \\
Proposed  &0.9412 & 4& 11 & $(936,936)$ & $(5,5)$ & $(2,2)$ & 436178 & $5.2860\cdot 10^{-3}$ & 5.4057  \\
 &0.9408 &5&  11 & $(1022,1022)$ & $(6,5)$ & $(2,2)$ & 522242 & $5.3341\cdot 10^{-3}$ & 5.3970  \\
  \hline
\cite[Sec. IV-C]{Smith12}$^{**}$ &0.9372 & 2& 10 & $(510,512)$ & $(3,3)$ & $(1,1)$ & 261120 & $5.6304\cdot 10^{-3}$ & 5.3490  \\\hdashline
\multirow{4}{*}{Proposed} &0.9372 & 2& 11 & $(876,876)$ & $(5,5)$ & $(3,3)$ & 255792 & $5.6427\cdot 10^{-3}$ & 5.3465  \\
 &0.9372 & 4& 11 & $(876,876)$ & $(5,5)$ & $(3,3)$ & 255792 & $5.6481\cdot 10^{-3}$ & 5.3453  \\
 &0.9372 &2& 11 & $(972,952)$ & $(6,5)$ & $(4,4)$ & 231336 & $5.6430\cdot 10^{-3}$ & 5.3466  \\
 &0.9372 & 5& 11 & $(964,964)$ & $(6,5)$ & $(4,4)$ & 232324 & $5.6550\cdot 10^{-3}$ & 5.3438  \\
 \hline
\cite[Table II]{6787025} &0.9333 &2& 11 & $(825,825)$ & $(5,5)$ & $(1,1)$ & 680625 &  $5.9922\cdot 10^{-3}$ & 5.2920  \\\hdashline
\multirow{2}{*}{Proposed} &0.9333 &2 & 11 & $(990,990)$ & $(6,6)$ & $(2,2)$ & 490050 &  $6.0145\cdot 10^{-3}$ & 5.2873 \\
 &0.9333 &4 & 11 & $(990,990)$ & $(6,6)$ & $(2,2)$ & 490050 &  $6.0246\cdot 10^{-3}$ & 5.2852 \\
  \hline
\cite[Table I]{7537380}& 0.9167 & 2& 10 & $(360,360)$ &$(3,3)$  & $(1,1)$ & 129600 & $7.9921\cdot 10^{-3}$  & 5.0053 \\\hdashline
Proposed & 0.9167 & 4& 10 & $(480,480)$ &$(4,4)$  & $(2,2)$ & 115200 & $8.1697\cdot 10^{-3}$  & 4.9763 \\
\hline
 \cite{8345919}$^*$ & 0.8672 & 2& 8 & $(128,128)$ &$(2,2)$  & $(1,1)$ & 16384 & $1.4016\cdot 10^{-2}$  & 4.4446 \\\hdashline
Proposed & 0.8671 & 4& 9 & $(237,237)$ &$(4,3)$  & $(3,3)$ & 18732 & $1.4288\cdot 10^{-2}$  & 4.4151 \\
 \hline
 \cite[Table I]{8856224}$^*$ & 0.8333 & 2& 9 & $(114,114)$ &$(2,2)$  & $(1,1)$ & 12996 & $1.5736\cdot 10^{-2}$  & 4.4345 \\\hdashline
\multirow{2}{*}{Proposed} & 0.8333 & 4& 9 & $(216,216)$ &$(4,4)$  & $(4,4)$ & 11664 & $1.8155\cdot 10^{-2}$  & 4.1987 \\
& 0.8340 & 5& 9 & $(244,244)$ &$(5,4)$  & $(4,4)$ & 14884 & $1.8145\cdot 10^{-2}$  & 4.1961 \\
 \hline
\end{tabular}
\begin{tablenotes}
      \scriptsize
      \item $^*$ BCH component codes extended by 1 parity bit, $^{**}$ BCH component codes extended by 2 parity bits.
    \end{tablenotes}
  \end{threeparttable}
\end{table*}

From Table \ref{table2}, it can be observed that the proposed codes achieve a larger threshold than the benchmark staircase codes for the same or similar rates and with comparable block sizes. The threshold gain becomes larger if the conventional staircase codes are with a small $t$, e.g., $t\leq 3$. More importantly, the actual coding gain of the proposed codes over staircase codes under iBDD can be larger than the corresponding threshold gain. This is because the thresholds gain is based on density evolution, where miscorrection-free iBDD is assumed \cite{7890389}. For the staircase codes with a small $t$, the error performance under iBDD will degrade due to miscorrection if their BCH component codes do not have any extended parity bits. In contrast, the proposed codes employ BCH component codes with larger $(t_1,t_2)$ such that the miscorrection probability can be greatly reduced. As a result, the actual coding gain of the proposed codes over staircase codes with a small $t$ under iBDD is larger than the threshold gain based on density evolution. Nevertheless, the threshold gain still provide insights into designing good codes with better waterfall performance.

\section{Error Floor Analysis}\label{sec:error_floor_analysis}
The error floor performance of the class of staircase codes is affected by stall patterns, which are referred to as a set of errors in the code block that cannot be corrected with iterative decoding as the number of iterations $\ell  \rightarrow \infty$. To determine the BER due to stall patterns, we consider a fixed code block $\boldsymbol{B}_i$ and the error bits of stall patterns including positions in $\boldsymbol{B}_i$ and possibly additional positions in $\boldsymbol{B}_{i+1},\ldots$ but not in $\boldsymbol{B}_{i-1}$. The BER of the error floor is dominated by the occurrence probability of the stall patterns with the smallest size \cite{Smith12,7537380}. Consider a BSC with crossover probability $p$. The BER can be approximated by using the union bound technique following \cite{Smith12}
\begin{align}\label{eq:error+_floor_ber}
\mathsf{BER}_{\mathsf{floor}}&\approx \frac{s_{\min}A_{\min}p^{s_{\min}}}{\frac{m_1m_2}{\min\{q_1,q_2\}}},
\end{align}
where $A_{\min}$ is the multiplicity of minimum stall patterns, and $s_{\min}$ is the number of error bits of a minimum stall pattern. The denominator $\frac{m_1m_2}{\min\{q_1,q_2\}}$ is the size of the code block in which a minimum stall pattern occurs. Since the stall patterns and the error floor behave completely different for different coupling widths, we analyze each term in \eqref{eq:error+_floor_ber} separately for different coupling widths. The analysis will be used to justify our choice of $q_1, q_2$, and $w$.

\subsection{Error Floor with $w=2$}\label{sec:error_floor_w2}

\subsubsection{Minimum Stall Pattern Analysis}\label{sec:error_floor_w2_smin}
We first assume that a stall patterns only appears in the received blocks $\boldsymbol{Y}_i$ and $\boldsymbol{Y}^{\pi}_i$ as it allows us to easily determine $s_{\min}$. We denote by $\boldsymbol{S}_i$ the stall pattern matrix associated with $\boldsymbol{Y}_i$ such that $\boldsymbol{Y}_i = \boldsymbol{B}_i+\boldsymbol{S}_i$. In other words, the position of each non-zero element in $\boldsymbol{S}_i$ corresponds to the position of an error bit in $\boldsymbol{Y}_i$. Likewise, the stall pattern matrix associated with $\boldsymbol{Y}^{\pi}_i$ is denoted by $\boldsymbol{S}^{\pi}_i$, which is obtained from $\boldsymbol{S}_i$ by following the transformation in \eqref{eq:int1}. We then have the following theorem for minimum stall patterns.
\begin{theorem}\label{prop:w2_smin}
Consider a SR-staircase codes with parameters $(t_1,t_2)$, $(q_1,q_2)$, and $w=2$. The exact number of the error bits of the minimum stall pattern is
\begin{align}\label{eq:smin_w2}
s_{\min} =& \min\bigg\{\max\left\{\left\lceil\frac{t_2+1}{q_1} \right\rceil  (t_1+1),\left\lceil\frac{t_1+1}{q_1} \right\rceil  (t_2+1)\right\}, \nonumber \\
&\max\left\{ \left\lceil\frac{t_1+1}{q_2} \right\rceil  (t_2+1),\left\lceil\frac{t_2+1}{q_2} \right\rceil  (t_1+1) \right\}\bigg\}.
\end{align}
\end{theorem}
\begin{IEEEproof}
See Appendix \ref{app2}.
\end{IEEEproof}
We use Example \ref{example3} to illustrate the idea of Theorem \ref{prop:w2_smin}.
\begin{figure}[t!]
	\centering
\includegraphics[width=3.1in,clip,keepaspectratio]{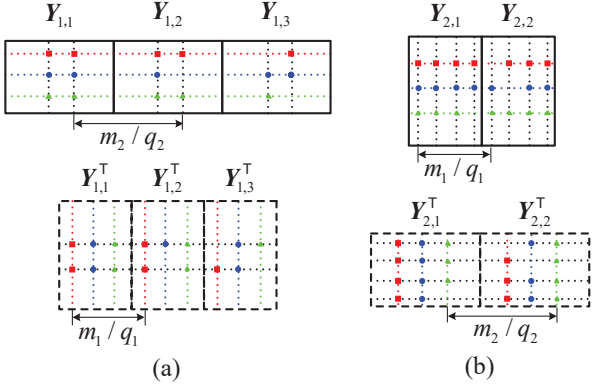}
\caption{Consider a SR-staircase code with $w=2$, $(t_1,t_2) = (6,4)$ and $(q_1,q_2) = (2,3)$. A stall pattern is formed in (a) block $\boldsymbol{Y}_1$, and (b) block $\boldsymbol{Y}_2$. The errors in the bit positions that belong to the same component codeword are represented by the same marker with the same color.}
\label{fig:s_staircase_stall1}
\end{figure}

\begin{example}\label{example3}
Consider a SR-staircase code with $w=2$, $(t_1,t_2) = (6,4)$ and $(q_1,q_2) = (2,3)$. The stall patterns formed in $\boldsymbol{Y}_1=[\boldsymbol{Y}_{1,1},\boldsymbol{Y}_{1,2},\boldsymbol{Y}_{1,3}]$ and $\boldsymbol{Y}_2=[\boldsymbol{Y}_{2,1},\boldsymbol{Y}_{2,2}]$ and their transformation in $\boldsymbol{Y}^{\pi}_1 =[\boldsymbol{Y}^{\mathsf{T}}_{1,1},\boldsymbol{Y}^{\mathsf{T}}_{1,2},\boldsymbol{Y}^{\mathsf{T}}_{1,3}]$ and $\boldsymbol{Y}^{\pi}_2 = [\boldsymbol{Y}^{\mathsf{T}}_{2,1},\boldsymbol{Y}^{\mathsf{T}}_{2,2}]$ are illustrated in Fig. \ref{fig:s_staircase_stall1}(a) and Fig. \ref{fig:s_staircase_stall1}(b), respectively. As shown in the top figure of Fig. \ref{fig:s_staircase_stall1}(a), the stall pattern formed in $\boldsymbol{Y}_1$ has a size of $3\times 5$. The transformation of this stall pattern in $\boldsymbol{Y}^{\pi}_1$ is shown in the bottom figure of Fig. \ref{fig:s_staircase_stall1}(a). On the other hand, a stall pattern with size $2\times 7$ can be formed in $\boldsymbol{Y}^{\pi}_1$ if we remove an error bit (represented by either the red, green, or blue marker) from the first erroneous row of $\boldsymbol{Y}^{\pi}_1$. However, this is equivalent to removing an error bit from either the first, third, or fifth erroneous column in $\boldsymbol{Y}_1$, leading to the correction of this stall pattern because one erroneous row in $\boldsymbol{Y}_1$ will have at most 4 errors and $t_2=4$. Therefore, only the $3\times 5$ stall pattern formed in $\boldsymbol{Y}_1$ is not correctable during the decoding of $[\boldsymbol{Y}^{\pi}_0,\boldsymbol{Y}_1]$ and $[\boldsymbol{Y}^{\pi}_1,\boldsymbol{Y}_2]$. Similarly, a stall pattern formed in $\boldsymbol{Y}_2$ has a minimum size of $3 \times 7$ as shown in the top figure of Fig. \ref{fig:s_staircase_stall1}(b), whereas its transformation in $\boldsymbol{Y}^{\pi}_2$ is illustrated in the bottom figure of Fig. \ref{fig:s_staircase_stall1}(b). As a result, we have $s_{\min} = 15$ as the stall pattern formed in $\boldsymbol{Y}_1$ has the smallest size. \demo
\end{example}

\subsubsection{Multiplicity Analysis}
To determine multiplicity $A_{\min}$, we consider that a minimum stall pattern can spread across $[\boldsymbol{Y}^{\pi}_{i}, \boldsymbol{Y}_{i+1}]$. From Theorem \ref{prop:w2_smin}, we see that whether a minimum stall pattern occurs in the block with even or odd index depends on $q_1$ and $q_2$. Hence, we can consider $q_1 \geq q_2$ without loss of generality. As a result, the minimum stall pattern occurs in $\boldsymbol{Y}_i,i\in2\mathbb{N}$.

By inspecting $s_{\min}$ in Theorem \ref{prop:w2_smin}, it can be seen that a minimum stall pattern affects exactly $\lceil\frac{t_1+1}{q_1}\rceil$ rows and at most $\lceil\frac{s_{\min}}{t_1+1}\rceil q_1$ columns in $[\boldsymbol{Y}^{\pi}_{i}, \boldsymbol{Y}_{i+1}]$. The intersections of these erroneous rows and columns form a rectangular array. We denote by $\mathcal{S}_{\mathsf{array}}$ and $\mathcal{S}_{\mathsf{stall}}$ the sets of error bit positions in the array and a minimum stall pattern, respectively, and define $\mathcal{Y}^{\pi}_i$ and $\mathcal{Y}_i$ to be the set of bit positions in $\boldsymbol{Y}^{\pi}_i$ and $\boldsymbol{Y}_i$, respectively. Clearly, we have $\mathcal{S}_{\mathsf{stall}} \subseteq \mathcal{S}_{\mathsf{array}} \subset (\mathcal{Y}^{\pi}_i \cup \mathcal{Y}_{i+1})$. The element in the $a$-th row and $b$-th column of matrix $[\boldsymbol{S}^{\pi}_{i},\boldsymbol{S}_{i+1}]$, i.e., $S_{i,i+1}(a,b)$, is 1 when $(a,b) \in \mathcal{S}_{\mathsf{stall}}$ and 0 otherwise. Then, $A_{\min}$ is the product of the number of ways to choose the positions of this array in $[\boldsymbol{Y}^{\pi}_{i},\boldsymbol{Y}_{i+1}]$ and the multiplicity of minimum stall patterns formed in the array. We denote by $A_{\mathsf{row}}$ and $A_{\mathsf{col}}$ the number of ways to choose row and column indices, respectively, for $\mathcal{S}_{\mathsf{array}}$. It is immediate that
\begin{align}\label{eq:arow}
A_{\mathsf{row}} = \binom{\frac{m_1}{q_1}}{\lceil \frac{t_1+1}{q_1} \rceil}.
\end{align}

To find $A_{\mathsf{col}}$, we further divide the aforementioned rectangular array into $\lceil\frac{s_{\min}}{t_1+1}\rceil$ sub-arrays of size $\lceil\frac{t_1+1}{q_1}\rceil \times q_1$ or $\lceil\frac{t_1+1}{q_1}\rceil \times q_2$. Hence, we have $\mathcal{S}_{\mathsf{array}} =  \bigcup_{j=1}^{\lceil\frac{s_{\min}}{t_1+1}\rceil}\mathcal{S}_{\mathsf{array},j}$, where $\mathcal{S}_{\mathsf{array},j}$ is the $j$-th sub-array. The rectangular array is divided such that the sub-array satisfies either $\mathcal{S}_{\mathsf{array},j}\subset \mathcal{Y}^{\pi}_i,\mathcal{S}_{\mathsf{array},j}\cap \mathcal{Y}_{i+1}=\varnothing$ or $\mathcal{S}_{\mathsf{array},j}\subset \mathcal{Y}_{i+1},\mathcal{S}_{\mathsf{array},j}\cap \mathcal{Y}^{\pi}_i=\varnothing$. In the former case, the sub-array is of size $\lceil\frac{t_1+1}{q_1}\rceil \times q_1$ and contains all possible positions of the error bits of an erroneous row vector in $\boldsymbol{Y}_i$. In the latter case, the sub-array is of size $\lceil\frac{t_1+1}{q_1}\rceil \times q_2$ and contains all possible positions of the error bits of an erroneous row vector in $\boldsymbol{Y}^{\pi}_{i+1}$. We denote by $(a,b)$ and $(a',b')$ a pair of position indices in $\mathcal{S}_{\mathsf{array},j}$, where $(a,b)\neq(a',b')$. Since all the bits of any erroneous row belong to the same component codeword, the column position indices of $\mathcal{S}_{\mathsf{array},j}$ satisfy $|b-b'| \in \{0,\frac{m_2}{q_2},\ldots,\frac{m_2 (q_1-1)}{q_2}\}$ when $\mathcal{S}_{\mathsf{array},j} \subset \mathcal{Y}^{\pi}_i$ and $|b-b'| \in \{0,\frac{m_1}{q_1},\ldots,\frac{m_1 (q_2-1)}{q_1}\}$ when $\mathcal{S}_{\mathsf{array},j} \subset \mathcal{Y}_{i+1}$. In other words, each sub-array always lies in the same column positions of each sub-block of $\boldsymbol{Y}^{\pi}_i$ or $\boldsymbol{Y}_{i+1}$. This means that given a column position of a sub-array $\mathcal{S}_{\mathsf{array},j}$, the rest of the column positions are deterministic. If there are $j$ sub-arrays in $\boldsymbol{Y}^{\pi}_{i}$, i.e., $\mathcal{S}_{\mathsf{array},1}\subset \mathcal{Y}^{\pi}_i,\ldots,\mathcal{S}_{\mathsf{array},j} \subset \mathcal{Y}^{\pi}_i$, then there are $\binom{\frac{m_2}{q_2}}{j}$ ways to choose all column indices for those $j$ sub-arrays. Similar arguments also apply to choosing the column indices for the other $\lceil\frac{s_{\min}}{t_1+1}\rceil - j$ sub-arrays in $\boldsymbol{Y}_{i+1}$. As a result, we obtain the multiplicity of the column indices for the rectangular error array as
\begin{align}\label{eq:acol}
A_{\mathsf{col}}
=& \binom{\frac{m_2}{q_2}}{\lceil\frac{s_{\min}}{t_1+1}\rceil}+\mathbbm{1}\left\{q_2 > \frac{t_1}{\lceil \frac{t_1+1}{q_1}\rceil}\right\} \times \nonumber \\
&\sum\nolimits_{j=1}^{\lceil\frac{s_{\min}}{t_1+1}\rceil-1}\binom{\frac{m_2}{q_2}}{j}\binom{\frac{m_2}{q_2}}{(\lceil\frac{s_{\min}}{t_1+1}\rceil-j)},
\end{align}
where the indicator function gives the condition that only the case $\mathcal{S}_{\mathsf{array}} \subset \mathcal{Y}^{\pi}_i,\mathcal{S}_{\mathsf{array}} \cap \mathcal{Y}_{i+1}= \varnothing$ is possible. The reasons are as follows. For any sub-array $\mathcal{S}_{\mathsf{array},j} \subset \mathcal{Y}_{i+1}$, we know that its size is $\lceil\frac{t_1+1}{q_1}\rceil \times q_2$. Since this sub-array contains all the possible positions of the error bits of an erroneous row in $\boldsymbol{Y}^{\pi}_{i+1}$, the erroneous row has at most $\lceil\frac{t_1+1}{q_1}\rceil q_2$ errors. This error vector is correctable by $\mathcal{C}_1$ if $\lceil\frac{t_1+1}{q_1}\rceil q_2 \leq t_1$.

Example \ref{exam:4} illustrates the relationship between a minimum stall pattern and its associated error array and sub-arrays.

\begin{example}\label{exam:4}
Consider the SR-staircase code in Example \ref{example3} again. In the bottom figure of Fig. \ref{fig:s_staircase_stall1}(a), a minimum stall pattern with size $s_{\min}=15$ is inside a $2 \times 9$ array in $\boldsymbol{Y}^{\pi}_1$. All the error bits of each erroneous row vector in $\boldsymbol{Y}_1$ shown in the top figure of Fig. \ref{fig:s_staircase_stall1}(a) are inside a $2 \times 3$ sub-array with its column positions marked by the dash lines with the same color in $\boldsymbol{Y}^{\pi}_1$ in the bottom figure. In the bottom figure of Fig. \ref{fig:s_staircase_stall1}(b), a (non-minimum) stall pattern formed in $\boldsymbol{Y}^{\pi}_2$ is inside a $4 \times 6$ array. All the error bits of each erroneous error vector in $\boldsymbol{Y}_2$ shown in the top figure in Fig. \ref{fig:s_staircase_stall1}(b) are inside a $4 \times 2$ sub-array in $\boldsymbol{Y}^{\pi}_2$.
Note that the $2 \times 9$ array in $\boldsymbol{Y}^{\pi}_1$ cannot spread into $\boldsymbol{Y}_2$. If any of its three $2\times 3$ sub-arrays is formed in $\boldsymbol{Y}_2$, then this sub-array will become an erroneous row with at most 4 error bits in $\boldsymbol{Y}^{\pi}_2$, which is correctable by $\mathcal{C}_2$. Hence, the minimum stall pattern can only be formed in $\boldsymbol{Y}^{\pi}_1$ rather than $[\boldsymbol{Y}^{\pi}_1,\boldsymbol{Y}_2]$.
\demo
\end{example}

It then remains to determine the multiplicity of minimum stall patterns formed in the error array. Following \eqref{eq:acol}, consider that there are $j$ sub-arrays in $\boldsymbol{Y}^{\pi}_i$ and the resultant error array is with size $\lceil\frac{t_1+1}{q_1}\rceil \times (jq_1+(\lceil\frac{s_{\min}}{t_1+1}\rceil-j)q_2)$. Next, we use an $\lceil\frac{t_1+1}{q_1}\rceil \times \lceil\frac{s_{\min}}{t_1+1}\rceil$ integer matrix $\boldsymbol{A}_{\mathsf{array},j}$ to represent the error number assignment which assigns the error bits of a minimum stall pattern to the error array with $j$ sub-arrays contained in $\boldsymbol{Y}^{\pi}_i$ and $\lceil\frac{s_{\min}}{t_1+1}\rceil-j$ sub-arrays contained in $\boldsymbol{Y}_{i+1}$. Its entry $A_{\mathsf{array},j}(i_1,i_2)$ with $i_1 \in \left[\lceil\frac{t_1+1}{q_1}\rceil\right]$ and $i_2 \in \left[\lceil\frac{s_{\min}}{t_1+1}\rceil\right]$, represents the number of errors in the $i_1$-th row of the $i_2$-th sub-array. More importantly, $A_{\mathsf{array},j}(i_1,i_2)$ must satisfy all conditions below
\begin{align}
 q_1 \geq& A_{\mathsf{array},j}(i_1,i_2) \geq t_1+1-\left(\left\lceil \frac{t_1+1}{q_1} \right\rceil-1\right)q_1, \nonumber \\
&\forall i_2 \in [j], \label{eq:relationship1}\\
q_2\geq& A_{\mathsf{array},j}(i_1,i_2) \geq t_1+1-\left(\left\lceil \frac{t_1+1}{q_2} \right\rceil-1\right)q_2 , \nonumber \\
&\forall i_2 \in \left[\left\lceil\frac{s_{\min}}{t_1+1}\right\rceil\right]\setminus[j],\label{eq:relationship2}\\
\sum\nolimits_{i_1=1}^ {\lceil\frac{t_1+1}{q_1}\rceil}& A_{\mathsf{array},j}(i_1,i_2)  \geq t_2+1, \forall  i_2 \in \left[\left\lceil\frac{s_{\min}}{t_1+1}\right\rceil\right],\label{eq:relationship3}\\
\sum\nolimits_{i_2=1}^{ \lceil\frac{s_{\min}}{t_1+1}\rceil}& A_{\mathsf{array},j}(i_1,i_2)  \geq t_1+1, \forall i_1 \in \left[\left\lceil\frac{t_1+1}{q_1}\right\rceil\right],\label{eq:relationship4}\\
\sum\nolimits_{i_1=1}^{ \lceil\frac{t_1+1}{q_1}\rceil}&\sum\nolimits_{i_2=1}^{ \lceil\frac{s_{\min}}{t_1+1}\rceil} A_{\mathsf{array},j}(i_1,i_2) = s_{\min}, \label{eq:relationship5}
\end{align}
where \eqref{eq:relationship1} and \eqref{eq:relationship2} give the ranges for the number of errors in each row of the $i_2$-th sub-array in $\boldsymbol{Y}^{\pi}_{i}$ and $\boldsymbol{Y}_{i+1}$, respectively, \eqref{eq:relationship3} enforces the constraint that each row of a minimum stall pattern must has at least $t_2+1$ errors, \eqref{eq:relationship4} enforces the constraint that the total number of errors contained by each sub-array must be at least $t_1+1$, and finally \eqref{eq:relationship5} gives the constraint on the total number of errors of a minimum stall pattern.
\begin{example}
In the bottom figure of Fig. \ref{fig:s_staircase_stall1}(a), the error number assignment of a minimum stall pattern to the $2 \times 9$ array in $\boldsymbol{Y}^{\pi}_1$ is $\boldsymbol{A}_{\mathsf{array},3}=\begin{bmatrix}
2 &3 &3 \\
3 &2 &2\\
\end{bmatrix}$.
Here, $j=3$ because all the three $2\times 3$ sub-arrays are in $\boldsymbol{Y}^{\pi}_1$. Moreover, the entries of the first to third columns in $\boldsymbol{A}_{\mathsf{array},3}$ correspond to the number of row errors in the sub-arrays marked with red, blue and green, respectively, in the bottom figure of Fig. \ref{fig:s_staircase_stall1}(a).
\demo
\end{example}

In addition to the error number assignment, we also need to determine the error position assignment. For the $i_1$-th row of the $i_2$-th sub-array, there are either $\binom{q_1}{A_{\mathsf{array},j}(i_1,i_2)}$ or $\binom{q_2}{A_{\mathsf{array},j}(i_1,i_2)}$ ways to assign $A_{\mathsf{array},j}(i_1,i_2)$ errors for this sub-array contained in either $\boldsymbol{Y}^{\pi}_i$ or $\boldsymbol{Y}_{i+1}$. The assignment for each entry in $\boldsymbol{A}_{\mathsf{array},j}$ is independent. Thus, given $\boldsymbol{A}_{\mathsf{array},j}$, the combinations of all row error assignments has the form of either $\prod_{i_1}\prod_{i_2}\binom{q_1}{A_{\mathsf{array},j}(i_1,i_2)}$ or $\prod_{i_1}\prod_{i_2}\binom{q_2}{A_{\mathsf{array},j}(i_1,i_2)}$.

Finally, with \eqref{eq:arow} and \eqref{eq:acol} and the number of combinations of minimum stall patterns formed in the error array, the multiplicity $A_{\min}$ for the case $q_1 \geq q_2$ is obtained as
\begin{figure*}[t]
\begin{align}\label{eq:Amin}
&A_{\min}=\binom{\frac{m_1}{q_1}}{\lceil \frac{t_1+1}{q_1} \rceil}\left(\binom{\frac{m_2}{q_2}}{\lceil\frac{s_{\min}}{t_1+1}\rceil}\sum_{\boldsymbol{A}_{\mathsf{array},\lceil\frac{s_{\min}}{t_1+1}\rceil}}\prod\nolimits_{i_1=1}^{ \lceil\frac{t_1+1}{q_1}\rceil}\prod\nolimits_{i_2=1}^{ \lceil\frac{s_{\min}}{t_1+1}\rceil}\binom{q_1}{A_{\mathsf{array},\lceil\frac{s_{\min}}{t_1+1}\rceil}(i_1,i_2)}+\mathbbm{1}\left\{q_2 > \frac{t_1}{\lceil \frac{t_1+1}{q_1}\rceil}\right\} \right.\nonumber \\
&\times\left.\sum\nolimits_{j=1}^{\lceil\frac{s_{\min}}{t_1+1}\rceil-1}\binom{\frac{m_2}{q_2}}{j}\binom{\frac{m_2}{q_2}}{(\lceil\frac{s_{\min}}{t_1+1}\rceil-j)}
 \left(\sum_{\boldsymbol{A}_{\mathsf{array},j}}\prod\nolimits_{i_1=1}^{ \lceil\frac{t_1+1}{q_1}\rceil}\prod\nolimits_{i_2=1}^{ j}\binom{q_1}{A_{\mathsf{array},j}(i_1,i_2)}\prod\nolimits_{i'_2=j+1}^{ \lceil\frac{s_{\min}}{t_1+1}\rceil}\binom{q_2}{A_{\mathsf{array},j}(i_1,i'_2)}\right)\right),
\end{align}
\hrule
\end{figure*}
where the summation over $\boldsymbol{A}_{\mathsf{array},j}$ takes all the possible non-identical $\boldsymbol{A}_{\mathsf{array},j}$ with each of its entry satisfying \eqref{eq:relationship1}-\eqref{eq:relationship5}. Finding the number of such matrices is closely related to the problem of matrices with prescribed row and column sums \cite{BARVINOK2012820}.

For the case of $q_2 \geq q_1$, the multiplicity $A_{\min}$ can be directly obtained from \eqref{eq:Amin} by swapping the argument between $m_1$ and $m_2$, $q_1$ and $q_2$, as well as $t_1$ and $t_2$.

\subsubsection{Code Block Index and Size}
We know that the minimum stall pattern occurs in $\boldsymbol{Y}_i$ for $i\in 2\mathbb{N}$ when $q_1\geq q_2$ and $i\in 2\mathbb{N}-1$ when $q_1\geq q_2$. Hence, the block which has the minimum stall pattern, contains $\frac{m_1m_2}{\min\{q_1,q_2\}}$ bits.

\begin{remark}\label{remark2}
Based on Theorem \ref{prop:w2_smin}, it is desirable to have $\max\{q_1,q_2\} \leq \min\{t_1,t_2\}$ when $w=2$ to ensure that any minimum stall pattern will not become a one-dimensional vector whose $s_{\min}$ becomes very small. Although the size of a minimum stall pattern for SR-staircase codes is smaller than that for the conventional staircase codes when both codes are with the same $(t_1,t_2)$, the proposed codes can still achieve a better error floor due to much smaller multiplicity $A_{\min}$ and the use of component codes with larger $(t_1,t_2)$. In addition, we note that the error floor can be improved by using some post-processing techniques proposed for the conventional staircase codes, e.g., \cite[Sec. V-A]{8425763}. For example, the simplest way is to flip the aforementioned error array that contains a minimum stall pattern, such that the residue errors will be corrected by iBDD. The error floor after post-processing will be studied in our future work.
\demo
\end{remark}

\subsection{Error Floor with $w>2$}
For a large coupling width, we need to set $m_1 = m_2 \triangleq m$ and $q_1 = q_2 \triangleq q$ according to Sec. \ref{subsec:enc_w3}. Moreover, we are particularly interested in the case of $w \geq q+1$ since this choice allows the proposed codes to achieve the largest decoding threshold as discussed at the end of Sec. \ref{sec:DE_result}. In the interest of space, we consider $w \geq q+1$ in the subsequent analysis.

\subsubsection{Minimum Stall Pattern Analysis}

Following Sec. \ref{sec:error_floor_w2_smin}, we use $\boldsymbol{S}_i$ and $\boldsymbol{S}^{\pi}_i$ to represent the stall pattern matrices associated with $\boldsymbol{Y}_i$ and $\boldsymbol{Y}^{\pi}_i$, respectively. For notation simplicity, we define the stall pattern matrix associated with the coupling sub-blocks in \eqref{enc:large_w}-\eqref{enc:large_w1} as $[\boldsymbol{S}^{\pi}_{i-l+1,l}]_{l=1}^{w-1}\triangleq[\boldsymbol{S}^{\pi}_{i,1},\boldsymbol{S}^{\pi}_{i+1,2},\ldots,\boldsymbol{S}^{\pi}_{i-w+2,w-1}]$. Obtaining the exact analytical expression for $s_{\min}$ is difficult as it varies with $w$. Alternatively, we derive a lower bound on $s_{\min}$, which will provide insights into the upper bound on the BER of the error floor.
\begin{theorem}\label{prop:largew}
Consider a SR-staircase code with parameters $(t_1,t_2)$, $m_1 = m_2 \triangleq m$, $q_1 = q_2 \triangleq q$, and $w \geq q+1$. The error number of the  minimum stall pattern is lower bounded by
\begin{align}\label{eq:smin_w_large}
s_{\min}\geq  \frac{(\min\{t_1,t_2\}+1)(\min\{t_1,t_2\}+2)}{2}.
\end{align}
\end{theorem}
\begin{IEEEproof}
See Appendix \ref{app3}
\end{IEEEproof}

Based on Theorem \eqref{prop:largew}, we have the following useful lemma.

\begin{lemma}\label{lem:stall_w3}
Consider the SR-staircase code in Theorem \eqref{prop:largew} with $w\geq q+1$ and assume $t_1 \neq t_2$. If $(q,w,t_1,t_2)$ further satisfy one of the following conditions: 1) $ \min \{t_1,t_2\} \geq q$; 2) $\min\{t_1,t_2\}+1 \leq q $ and $ w \leq 2(\min \{t_1,t_2\}+1)$, $s_{\min}$ is strictly larger than the lower bound in \eqref{eq:smin_w_large}.
\end{lemma}
\begin{IEEEproof}
See Appendix \ref{app4}.
\end{IEEEproof}
Corollary \ref{corollary1} follows immediately from Theorem \eqref{prop:largew} and Lemma \ref{lem:stall_w3} and their proofs in Appendices \ref{app3}-\ref{app4}.
\begin{corollary}\label{corollary1}
For the SR-staircase code in Theorem \ref{prop:largew} with $w\geq q+1$, $s_{\min}$ achieves its lower bound in \eqref{eq:smin_w_large} if and only if $w \geq  (\mathbbm{1}\{t_1\neq t_2\}+1) (\min\{t_1,t_2\}+1)+1$ and $ q \geq \min\{t_1,t_2\}+1$.
\end{corollary}

\begin{remark}
Notice that all of our designs in Table \ref{table2} satisfy $|t_1-t_2| \in \{0,1\}$ because these designs achieve a better threshold than those with $|t_1-t_2| >1$. Under this condition, the lower bound of $s_{\min}$ in Theorem \ref{prop:largew} is larger than the exact $s_{\min}$ for $w=2$, $q_1 \geq 2$ and $q_2 \geq 2$ in Theorem \ref{prop:w2_smin}. Hence, the error floor can be improved by increasing $w$. In addition, Lemma \ref{lem:stall_w3} shows that if both $q$ and $w$ are not too large, the size of the minimum stall pattern can become larger. In fact, Tables \ref{thres1}-\ref{table2} already show that a moderate value of $q$ and $m$ suffice to achieve the best decoding threshold. Hence, a proper choice of $(q,w,t_1,t_2)$ would lead to a better trade-off between waterfall and error floor for SR-staircase codes.

\end{remark}

\subsubsection{Multiplicity Analysis}
We find $A_{\min}$ by assuming that $s_{\min}$ achieves its lower bound. Hence, the code parameters satisfies the conditions in Corollary \ref{corollary1}.

To begin with, we assign a row of $\min\{t_1,t_2\}+1$ errors to $\boldsymbol{B}_i$ such that the conditions of \eqref{eq62a} and \eqref{eq62} in Appendix \ref{app4} are satisfied. Consider an erroneous row with index $r_\mathsf{c}$ in $\left[[\boldsymbol{Y}^{\pi}_{i-l+\bar{z},l}]_{l=1}^{w-1},\boldsymbol{Y}_{i+\bar{z}}\right]$, where $\bar{z} =\lceil\frac{q  r_\mathsf{c}}{m} \rceil$ and $\bar{z} \in [w-1]\cap (2\mathbb{N})$ by \eqref{eq:s_con_odd} in Appendix \ref{app4}. From \eqref{eq:s_con_odd_0}-\eqref{eq:rc1_con2} in Appendix \ref{app4}, we know that the number of errors of each affected row is deterministic. As for the positions of those error bits, it can be seen that the column position of each error bit in $[\boldsymbol{Y}^{\pi}_{i-l+\bar{z},l}]_{l=1}^{w-1}$ is determined by the row position of that bit in the previous received block. Meanwhile, the row position of each error bit in $[\boldsymbol{Y}^{\pi}_{i-l+\bar{z},l}]_{l=1}^{w-1}$ must be the same as that for the erroneous row in $\boldsymbol{Y}_{i+\bar{z}}$, which also determines the column position of that bit in the succeeding coupled blocks. In other words, once a row of $\min\{t_1,t_2\}+1$ errors are assigned to $\boldsymbol{B}_i$, the row and column positions of the rest of the error bits are determined. Therefore, the multiplicity is
\begin{align}
A_{\min} &= \binom{\lfloor\frac{w-1}{\mathbbm{1}\{t_1\neq t_2\}+1}\rfloor}{\min\{t_1,t_2\}+1}\binom{\frac{m}{w-1}}{1}\left(\binom{\frac{m}{q}}{1}\right)^{\min\{t_1,t_2\}+1} \label{eq:Amin_w3} \\
&\geq \frac{m^{\min\{t_1,t_2\}+2}}{(w-1)q^{\min\{t_1,t_2\}+1}}, \label{eq:Amin_w3a}
\end{align}
where \eqref{eq:Amin_w3a} holds for $ w  = (\mathbbm{1}\{t_1\neq t_2\}+1) (\min\{t_1,t_2\}+1)+1$. Plugging \eqref{eq:Amin_w3} and \eqref{eq:smin_w_large} into \eqref{eq:error+_floor_ber} gives the estimation of the error floor.

\begin{figure}[t!]
	\centering
\includegraphics[width=3.1in,clip,keepaspectratio]{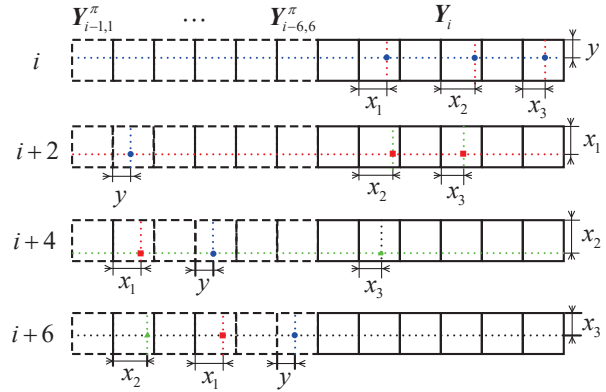}
\caption{Consider a SR-staircase code with $w=7$, $(t_1,t_2)=(2,3)$ and $q=6$. A minimum stall pattern with $s_{\min}=6$ is formed in blocks $\boldsymbol{Y}_i,\boldsymbol{Y}_{i+2}$ and $\boldsymbol{Y}_{i+4}$ as well as their corresponding coupled sub-blocks. The position of each error bit is also shown.}
\label{fig:s_staircase_stall1_w3}
\end{figure}

\begin{example}
Consider a SR-staircase code with $(t_1,t_2,q,w)=(2,3,6,7)$. Fig. \ref{fig:s_staircase_stall1_w3} shows a minimum stall pattern with $s_{\min}=6$ formed in $\left[[\boldsymbol{Y}^{\pi}_{i-l+\bar{z},l}]_{l=1}^{6},\boldsymbol{Y}_{i+\bar{z}}\right]$ for $\bar{z} = 0,2,4,6$. As shown in Fig. \ref{fig:s_staircase_stall1_w3}, given the row and column positions of each error bit in $\boldsymbol{Y}_i$, the positions of the rest of the error bits are deterministic. \demo
\end{example}

If $(t_1,t_2,q,w)$ satisfy the conditions in Lemma \ref{lem:stall_w3}, the minimum stall pattern size is strictly larger than \eqref{eq:smin_w_large}. Since it is difficult to find the exact minimum stall pattern size and its multiplicity in this case, we can use \eqref{eq:smin_w_large} and \eqref{eq:Amin_w3a} to obtain an upper bound of its true error floor.

\section{Numerical Results}\label{sec:sim}

We evaluate the performance of SR-staircase codes over the AWGN channel. A maximum of ten decoding iterations were performed over a decoding window. It should be noted that all BCH component codes used in our designs do not have any extended parity bits.

\begin{figure}[t!]
	\centering
\includegraphics[width=3.2in,clip,keepaspectratio]{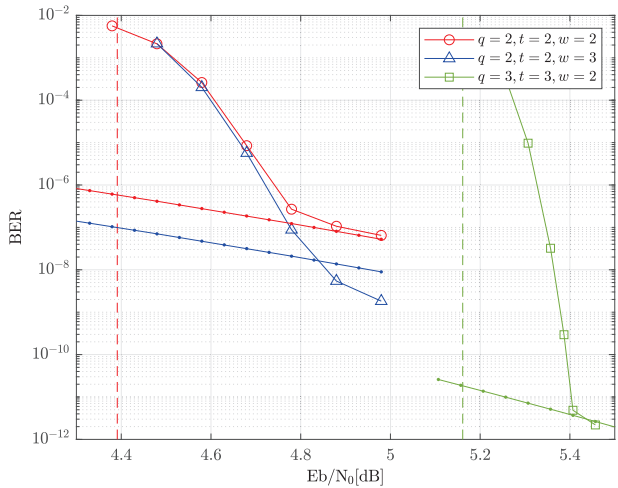}
\caption{BER of SR-staircase codes with their decoding threshold (dash line) and error floor estimation (with marker `$\cdot$').}
\label{fig:BER_floor_v1}
\end{figure}

We first use simulation results to validate our theoretical analysis by assuming miscorrection-free iBDD. We construct three SR-staircase codes with parameters $(m,\nu,t,q,w)=(126,8,2,2,2)$, $(126,8,2,2,3)$, and $(441,9,3,3,2)$, respectively. The decoding window size is set to $W=7$. The simulated BER, decoding threshold and the estimated error floor $\mathsf{BER}_{\mathsf{floor}}$ are shown in Fig. \ref{fig:BER_floor_v1}. For the SR-staircase codes with $w=2$, their simulated error floor BER matches closely to $\mathsf{BER}_{\mathsf{floor}}$ based on Theorem \ref{prop:w2_smin} and \eqref{eq:Amin}. Clearly, increasing $w$ leads to a lower error floor. It is also interesting to note that the code with $w=3$ achieves a lower error floor than its estimated error floor $\mathsf{BER}_{\mathsf{floor}}$. This is because the code parameters $(t,q,w)=(2,2,3)$ satisfy the conditions in Lemma \ref{lem:stall_w3} such that the size of the minimum stall pattern is strictly larger than that in Theorem \ref{prop:largew}. Consequently, the $\mathsf{BER}_{\mathsf{floor}}$ based on \ref{eq:smin_w_large} and \eqref{eq:Amin_w3a} can only serve as an upper bound of the true error floor. Observe that the simulated waterfall performance for all the codes is also in agreement with the derived decoding threshold (the threshold curves for the codes with $t=2$ and $w\in \{2,3\}$ are overlapped). Therefore, both DE and error floor analysis can be used to effectively predict the simulated performance if the probability of miscorrection is low, which is the case in our subsequent design with a large $t$.

\begin{figure}[t!]
	\centering
\includegraphics[width=3.2in,clip,keepaspectratio]{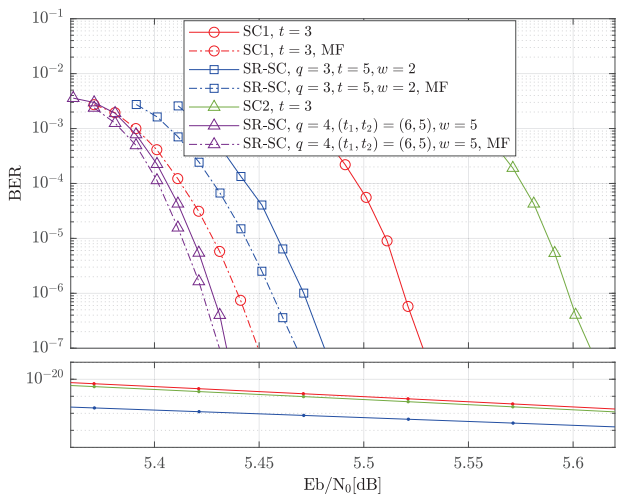}
\caption{Simulation results for SR-staircase codes and the staircase codes from \cite{Smith12}.}
\label{fig:BER_v1}
\end{figure}

Next, we compare the designed SR-staircase codes with the conventional staircase codes. For SR-staircase codes (labeled as ``SR-SC''), we consider two designs from Table \ref{table2}, whose parameters are $(m,\nu,t,q,w)=(876,11,5,3,2)$, and $(m,\nu,t_1,t_2,q,w)=(964,11,6,5,4,5)$, respectively. We also consider two benchmark conventional staircase codes, where the first one (labeled as ``SC1'') has parameters $(m_1,m_2,\nu,t) = (510,512,10,3)$ and two parity bits extended for BCH component codes following \cite[Sec. IV-C]{Smith12} while the second one (labeled as ``SC2'') has parameters $(m,\nu,t) = (478,10,3)$ and no extended parity bits. Notice that the BCH component codes of SR-staircase codes have a larger minimum distance than those of staircase codes. Thus, the decoding complexity of SR-staircase codes is expected to be higher than that of the benchmark staircase codes. All the codes have rate 0.9372 and comparable code block size as shown in Table \ref{table2}. The decoding window size is set to $W=9$ for demonstration purposes. It can be reduced for achieving a lower decoding latency \cite{7296605} at the cost of slightly inferior waterfall performance for both types of codes. The BER under iBDD (solid lines), miscorrection-free iBDD (dashed lines, labeled as ``MF''), and the estimated error floor $\mathsf{BER}_{\mathsf{floor}}$ are shown in Fig. \ref{fig:BER_v1} (the $\mathsf{BER}_{\mathsf{floor}}$ of the SR-staircase code with $w=5$ is not shown in the figure as it is in the order of $10^{-33}$). Observe that SC2 under iBDD has the worst performance due to the highest probability of miscorrection. Even though SC1 uses two additional parity bits to reduce miscorrection probability, it still has a noticeable gap to its miscorrection-free performance. In contrast, all the proposed codes operate close to their miscorrection-free performance with iBDD and outperform the conventional staircase codes in terms of better waterfall and error floor performance. Most notably, the SR-staircase code with $w=5$ has the best performance among all the codes and achieves slightly better waterfall performance with iBDD than the convectional staircase code with miscorrection-free iBDD.

\section{Concluding Remarks}\label{sec:conclude}
We proposed SR-staircase codes, a new class of spatially coupled product codes. The proposed codes are derived from the conventional staircase codes and have a larger design space. The most appealing feature is that one can employ stronger BCH component codes to construct a SR-staircase code with a similar or the same rate and block size as staircase code. The decoding threshold and the error floor of SR-staircase codes were analyzed by using DE and the union bound technique, respectively. Both theoretical and simulation results demonstrate the superior performance of the proposed codes over staircase codes in terms of waterfall and error floor. In addition, it was shown that increasing the coupling width can further improve the performance.

For future works, it would be interesting to consider the design and analysis of the proposed SR-staircase codes with other component codes. Another worthwhile direction could be designing low-complexity concatenating coding schemes for soft-decision channels, where inner codes will use soft-decision decoding and SR-staircase codes under iBDD will be used as outer codes.

\appendices

\section{Proof of Theorem \ref{prop1}}\label{app1}

First, in order to satisfy the rate requirement, we have
\begin{align}\label{eq:req_rate}
R \geq R' \overset{\eqref{eq:rate}}{\Rightarrow} & 1-\frac{t\nu}{m}  \geq 1-\frac{t'\nu'}{m'}
\Rightarrow  m \geq \frac{t\nu m'}{t'\nu'}.
\end{align}

Then, to satisfy the BSC threshold requirement, we have
\begin{align}\label{eq:req_thres}
\bar{p} > \bar{p}' \overset{\eqref{eq:MP}}{\Rightarrow}&  \frac{\bar{M}}{m}>\frac{\bar{M}'}{m'}
\Rightarrow m  < \frac{\bar{M}}{\bar{M}'}m'.
\end{align}

The block size requirement leads to
\begin{align}\label{eq:req_block}
\frac{m^2}{q} \leq (m')^2  \Rightarrow m \leq \sqrt{q}m'.
\end{align}

Combining \eqref{eq:req_rate}-\eqref{eq:req_block} and the fact that $2m \leq 2^\nu -1$, the resultant SR-staircase code has $R \geq R'$, $\bar{p}>\bar{p}'$ and $\frac{m^2}{q} \leq (m')^2$ if $m$ satisfies
\begin{align}\label{eq:ine_m1}
\beta\left\lceil\frac{t\nu m'}{t'\nu'\beta} \right\rceil  \leq m <\min\left\{\sqrt{q}m',\frac{\bar{M}}{\bar{M}'}m',\frac{2^\nu -1}{2}\right\},
\end{align}
where the operation $\lceil.\rceil$ and $\beta\triangleq \text{LCM}(w-1,q)$ ensure that the left boundary point in \eqref{eq:ine_m1} is divisible by both $q$ and $w-1$. We therefore obtain the necessary condition \eqref{eq:req1} from \eqref{eq:ine_m1}.

\section{Proof of Theorem \ref{prop:w2_smin}}\label{app2}

First, we consider $i\in 2\mathbb{N}$. Define $\boldsymbol{s}_{i,r_1}$ to be the non-zero row of stall pattern matrix $\boldsymbol{S}_{i}$ with index $r_1$, $\boldsymbol{s}^{\pi}_{i,r_2}$ to be the non-zero row of the transformed stall pattern matrix $\boldsymbol{S}^{\pi}_{i}$ with index $r_2$, and $\mathcal{R}_1$ and $\mathcal{R}_2$ to be the collections of indices $r_1$ and $r_2$, respectively, where $\mathcal{R}_1 \subseteq [\frac{m_2}{q_2}]$ and $\mathcal{R}_2\subseteq [\frac{m_1}{q_1}]$. Since any stall pattern in $\boldsymbol{Y}_i$ must not be correctable during the decoding of $[\boldsymbol{Y}^{\pi}_{i-1},\boldsymbol{Y}_{i}]$ and $[\boldsymbol{Y}^{\pi}_{i},\boldsymbol{Y}_{i+1}]$, then each non-zero row of $\boldsymbol{S}_i$ and $\boldsymbol{S}^{\pi}_i$ must satisfy
\begin{align}
w_{\mathsf{H}}(\boldsymbol{s}_{i,r_1}) \geq& t_1+1, \nonumber\\
\forall r_1 \in& \mathcal{R}_1 \triangleq \{r_1|\boldsymbol{s}_{i,r_1}\in \boldsymbol{S}_i,\boldsymbol{s}_{i,r_1}\neq \boldsymbol{0}\}, \label{eq:sil1}\\
w_{\mathsf{H}}(\boldsymbol{s}^{\pi}_{i,r_2}) \geq& t_2+1, \nonumber\\
\forall r_2 \in& \mathcal{R}_2 \triangleq \{r_2|\boldsymbol{s}^{\pi}_{i,r_2}\in \boldsymbol{S}^{\pi}_i,\boldsymbol{s}^{\pi}_{i,r_2}\neq \boldsymbol{0}\},\label{eq:sipil2}\\
\sum_{r_1\in \mathcal{R}_1}w_{\mathsf{H}}(\boldsymbol{s}_{i,r_1}) =& \sum_{r_2\in \mathcal{R}_2}w_{\mathsf{H}}(\boldsymbol{s}^{\pi}_{i,r_2})\label{si_sipi_con}.
\end{align}

Recall that since $\boldsymbol{B}^{\pi}_i = [\boldsymbol{B}^{\pi}_{i,1},\ldots,\boldsymbol{B}^{\pi}_{i,q_1}]$, all bits in the same column position of every sub-block $\boldsymbol{B}^{\pi}_{i,l},l\in[q_1]$ belong to the same component codeword of $\mathcal{C}_1$. This means that for any non-zero row vector with no less than $t_1+1$ errors in $\boldsymbol{S}_i$, all these error bits occupy at least $\lceil\frac{t_1+1}{q_1} \rceil$ rows in $\boldsymbol{S}^{\pi}_i$ due to the transformation in \eqref{eq:int1} in Section \ref{subsec:enc}. Thus, the lower bounds on the required number of non-zero rows and error bits in $\boldsymbol{S}^{\pi}_i$ to form a stall pattern are
\begin{align}
|\mathcal{R}_2| \geq& \left\lceil\frac{t_1+1}{q_1} \right\rceil , \label{eq:spi_i_row_no}\\
 \Rightarrow  \sum_{r_2\in \mathcal{R}_2}w_{\mathsf{H}}(\boldsymbol{s}^{\pi}_{i,r_2}) \geq& |\mathcal{R}_2|\min_{r_2 \in\mathcal{R}_2}\left\{w_{\mathsf{H}}(\boldsymbol{s}^{\pi}_{i,r_2})\right\} \nonumber \\
 \geq&\left\lceil\frac{t_1+1}{q_1} \right\rceil \cdot (t_2+1)\label{eq:spi_i}.
\end{align}

We are left with determining the required minimum number of the error bits in $\boldsymbol{S}_i$ to form a stall pattern. To ensure that all the error bits of each erroneous row vector in $\boldsymbol{S}_i$ only occupy at most $\lceil\frac{t_1+1}{q_1} \rceil$ rows in $\boldsymbol{S}^{\pi}_i$ (otherwise, the size of the stall pattern in $\boldsymbol{S}^{\pi}_i$ would become larger), each non-zero row of $\boldsymbol{S}_i$ must satisfy
\begin{align}\label{eq:wh_si}
\left\lceil\frac{t_1+1}{q_1} \right\rceil\cdot q_1 \geq& w_{\mathsf{H}}(\boldsymbol{s}_{i,r_1}) \nonumber \\
\geq& t_1+1 , \forall r_1 \in \mathcal{R}_1.
\end{align}

Then, we obtain the minimum number of non-zero rows and error bits of $\boldsymbol{S}_i$
\begin{align}
|\mathcal{R}_1| \geq& \left\lceil\frac{\sum_{r_1\in \mathcal{R}_1}w_{\mathsf{H}}(\boldsymbol{s}_{i,r_1})}{\max\{w_{\mathsf{H}}(\boldsymbol{s}_{i,r_1})\}} \right\rceil  \nonumber\\
\overset{\eqref{si_sipi_con}}{=}&\left\lceil\frac{\sum_{r_2\in \mathcal{R}_2}w_{\mathsf{H}}(\boldsymbol{s}^{\pi}_{i,r_2})}{\max\{w_{\mathsf{H}}(\boldsymbol{s}_{i,r_1})\}} \right\rceil \nonumber \\ \overset{\eqref{eq:spi_i} \& \eqref{eq:wh_si} }{\geq}&
\left\lceil\frac{t_2+1}{q_1} \right\rceil, \nonumber \\
 \Rightarrow  \sum_{r_1\in \mathcal{R}_1}w_{\mathsf{H}}(\boldsymbol{s}_{i,r_1}) \geq& |\mathcal{R}_1|\min_{r_1 \in\mathcal{R}_1}\left\{w_{\mathsf{H}}(\boldsymbol{s}_{i,r_1})\right\}\nonumber \\
 \geq& \left\lceil\frac{t_2+1}{q_1} \right\rceil \cdot (t_1+1) \label{eq:si_min_size}.
\end{align}

To ensure that the conditions of \eqref{si_sipi_con}, \eqref{eq:spi_i} and \eqref{eq:si_min_size} are fulfilled simultaneously, the minimum number of error bits to form a stall pattern in $\boldsymbol{Y}_i$ with $i\in 2\mathbb{N}$ is obtained as
\begin{align}
s_{\min} =&  \max\left\{\min\left\{\sum_{r_1\in\mathcal{R}_1}w_{\mathsf{H}}(\boldsymbol{s}_{i,r_1})\right\},\right.\nonumber \\
&\left.\min\left\{\sum_{r_2\in \mathcal{R}_2}w_{\mathsf{H}}(\boldsymbol{s}^{\pi}_{i,r_2})\right\} \right\} \\
=& \max\left\{\left\lceil\frac{t_2+1}{q_1} \right\rceil(t_1+1),\left\lceil\frac{t_1+1}{q_1} \right\rceil (t_2+1)\right\}.\label{s_min_even}
\end{align}
The $s_{\min}$ for the case of $i\in 2\mathbb{N}+1$ can be easily obtained from \eqref{s_min_even} by swapping the subscripts between 1 and 2. By taking the minimum of $s_{\min}$ obtained for these two cases, the expression in \eqref{eq:smin_w2} of Theorem \ref{prop:w2_smin} follows.

\section{Proof of Theorem \ref{prop:largew}}\label{app3}

We first consider that the stall pattern spreads from block $\boldsymbol{Y}_i$ to $\boldsymbol{Y}_{i+1},\ldots$. Following \eqref{eq:sil1} in Appendix \ref{app2}, the weight of each non-zero row in $\boldsymbol{S}_i$ satisfies
\begin{align}\label{eq:w3con1}
w_{\mathsf{H}}(\boldsymbol{s}_{i,r_1}) \geq t_{\varphi(i)}+1, \forall r_1 \in \mathcal{R}_1\subseteq \left[\frac{m}{q}\right].
\end{align}

Due to coupling \eqref{enc:large_w1}, the errors in one erroneous row vector in $\boldsymbol{Y}_i$ will spread to some of the $w-1$ consecutive received blocks $\boldsymbol{Y}_{i+1},\ldots,\boldsymbol{Y}_{i+w-1}$ and affect at least $t_{\varphi(i)}+1$ rows, where each affected row is not correctable if a stall pattern is formed. This is because since $w \geq q+1$, the erroneous row of each decomposed sub-blocks of $\boldsymbol{Y}_{i}$ will become a column of errors in different received blocks due to coupling. Then, the corresponding stall pattern matrices satisfy
\begin{align}
&\boldsymbol{S}_{i}\neq \boldsymbol{0} \Rightarrow \boldsymbol{S}^{\pi}_{i} = [\boldsymbol{S}^{\pi}_{i,1},\ldots,\boldsymbol{S}^{\pi}_{i,w-1}]  \neq \boldsymbol{0}\label{eq:wL_con0} \\
 \Rightarrow& \left[\left([\boldsymbol{S}^{\pi}_{i-l+1,l}]_{l=1}^{w-1}\right)^\mathsf{T},\left([\boldsymbol{S}^{\pi}_{i-l+2,l}]_{l=1}^{w-1}\right)^\mathsf{T},\ldots,\right.\nonumber \\
 &\left.\left([\boldsymbol{S}^{\pi}_{i-l+w-1,l}]_{l=1}^{w-1}\right)^\mathsf{T}\right]^\mathsf{T} \neq \boldsymbol{0}\label{eq:47} \\
  \Rightarrow& \boldsymbol{S}^{\pi}_{\Sigma}\triangleq \left[\left([\boldsymbol{S}^{\pi}_{i-l+1,l}]_{l=1}^{w-1}\right)^\mathsf{T},\left([\boldsymbol{S}^{\pi}_{i-l+2,l}]_{l=1}^{w-1}\right)^\mathsf{T},\ldots,\right. \nonumber \\
  &\left.\left([\boldsymbol{S}^{\pi}_{i-l+\tau,l}]_{l=1}^{w-1}\right)^\mathsf{T}\right]^\mathsf{T} \neq \boldsymbol{0},\nonumber \\
  &\boldsymbol{S}_{\Sigma} \triangleq\left[\boldsymbol{S}^\mathsf{T}_{i+1},\ldots,\boldsymbol{S}^\mathsf{T}_{i+\tau}\right]^\mathsf{T} \neq \boldsymbol{0} ,\label{eq:wL_con1}
\end{align}
where \eqref{eq:47} follows by combining the stall pattern matrices associated with all the coupled sub-blocks, and \eqref{eq:wL_con1} follows by considering the worst case where the stall pattern spreads to $[[\boldsymbol{Y}^{\pi}_{i-l+\tau,l}]_{l=1}^{w-1},\boldsymbol{Y}_{i+\tau}]^\mathsf{T}$ for some $\tau \geq w-1$. For notation simplicity, we define the combined received block $[\boldsymbol{Y}^{\pi}_{\Sigma},\boldsymbol{Y}_{\Sigma}] \triangleq \left[[\boldsymbol{Y}^{\pi}_{i-l+1,l}]_{l=1}^{w-1},\boldsymbol{Y}_{i+1}]^\mathsf{T},\ldots,[[\boldsymbol{Y}^{\pi}_{i-l+\tau,l}]_{l=1}^{w-1},\boldsymbol{Y}_{i+\tau}]^\mathsf{T}\right]^\mathsf{T}$, such that all erroneous rows except those in $\boldsymbol{Y}_i$, are in this combined received block. Then, we obtain the following conditions on each non-zero row of the corresponding combined stall pattern matrix $[\boldsymbol{S}^{\pi}_{\Sigma},\boldsymbol{S}_{\Sigma}]$ with $\boldsymbol{S}^{\pi}_{\Sigma}$ and $\boldsymbol{S}_{\Sigma}$ defined in \eqref{eq:wL_con1}
\begin{align}\label{eq:w3con2}
&w_{\mathsf{H}}([\boldsymbol{s}^{\pi}_{r_\mathsf{c}},\boldsymbol{s}_{r_\mathsf{c}}]) \geq \min\{t_1,t_2\}+1, \nonumber \\
&\forall r_\mathsf{c} \in \mathcal{R}_\mathsf{c}\triangleq \left\{r_\mathsf{c}\Big|\boldsymbol{s}^{\pi}_{r_\mathsf{c}}\in\boldsymbol{S}^{\pi}_{\Sigma},\boldsymbol{s}_{r_\mathsf{c}}\in \boldsymbol{S}_{\Sigma},[\boldsymbol{s}^{\pi}_{r_\mathsf{c}},\boldsymbol{s}_{r_\mathsf{c}}]\neq \boldsymbol{0}\right\},
\end{align}
where $r_\mathsf{c}$ is the index of a non-zero row in $[\boldsymbol{S}^{\pi}_{\Sigma},\boldsymbol{S}_{\Sigma}]$ and $\mathcal{R}_\mathsf{c} \subseteq [\frac{\tau m}{q} ]$ denotes the corresponding set of indices. Then, the total number of affected rows in $[\boldsymbol{Y}^{\pi}_{\Sigma},\boldsymbol{Y}_{\Sigma}]$ is lower bounded by the minimum number of row errors in $\boldsymbol{Y}_i$ due to coupling and \eqref{eq:w3con1}
\begin{align}\label{eq:w3con3}
|\mathcal{R}_\mathsf{c}| \geq w_{\mathsf{H}}(\boldsymbol{s}_{i,r_1})\geq t_{\varphi(i)}+1.
\end{align}

With \eqref{eq:w3con2} and \eqref{eq:w3con3}, we lower bound the total number of error bits occur in $[\boldsymbol{Y}^{\pi}_{\Sigma},\boldsymbol{Y}_{\Sigma}]$
\begin{align}\label{eq:51a}
\sum_{r_\mathsf{c}\in\mathcal{R}_\mathsf{c}}w_{\mathsf{H}}([\boldsymbol{s}^{\pi}_{r_\mathsf{c}},\boldsymbol{s}_{r_\mathsf{c}}])
\geq& |\mathcal{R}_\mathsf{c}|(\min\{t_1,t_2\}+1) \nonumber \\
 \geq&(t_{\varphi(i)}+1)(\min\{t_1,t_2\}+1).
\end{align}

Finally, using the fact that the number of errors in each erroneous code block is equal to that of its transformation (e.g., the number of errors in $\boldsymbol{S}_i$ is the same as in $\boldsymbol{S}^{\pi}_{i}$), we obtain that
\begin{align}\label{eq:51}
s_{\min} = & \sum_{r_\mathsf{c}\in\mathcal{R}_\mathsf{c}}w_{\mathsf{H}}(\boldsymbol{s}^{\pi}_{r_\mathsf{c}})\nonumber \\
=& \sum_{r_\mathsf{c}\in\mathcal{R}_\mathsf{c}}w_{\mathsf{H}}(\boldsymbol{s}_{r_\mathsf{c}})+\sum_{r_1\in\mathcal{R}_1}w_{\mathsf{H}}(\boldsymbol{s}_{i,r_1}) \nonumber\\
=&\frac{\sum_{r_\mathsf{c}\in\mathcal{R}_\mathsf{c}}w_{\mathsf{H}}([\boldsymbol{s}^{\pi}_{r_\mathsf{c}},\boldsymbol{s}_{r_\mathsf{c}}])+\sum_{r_1\in\mathcal{R}_1}w_{\mathsf{H}}(\boldsymbol{s}_{i,r_1})}{2} \nonumber\\
\overset{\eqref{eq:w3con1} \&\eqref{eq:51a}}{\geq}&\frac{(t_{\varphi(i)}+1)(\min\{t_1,t_2\}+1)+(t_{\varphi(i)}+1)}{2} \nonumber\\
\geq&\frac{(\min\{t_1,t_2\}+2)(\min\{t_1,t_2\}+1)}{2}.
\end{align}

\section{Proof of Lemma \ref{lem:stall_w3}}\label{app4}
We use the notations and definitions from Appendix \ref{app3} and prove this lemma by contradiction. Consider $t_1 < t_2$ without loss of generality and assume that $s_{\min}$ achieves the lower bound in \eqref{eq:smin_w_large} in Theorem \ref{prop:largew} with $(t_1,t_2,q,w)$ satisfying the conditions in Lemma \ref{lem:stall_w3}, i.e.,
\begin{align}\label{eq:app4_contra}
s_{\min}=\frac{(t_1+1)(t_1+2)}{2}.
\end{align}

By \eqref{eq:w3con1}, \eqref{eq:w3con2} and \eqref{eq:w3con3} in Appendix \ref{app3}, the following conditions must hold simultaneously
\begin{align}
&w_{\mathsf{H}}(\boldsymbol{s}_{i,r_1}) = t_1+1, r_1\in \mathcal{R}_1,|\mathcal{R}_1|=1,\label{eq:w3con1_new}\\
&w_{\mathsf{H}}([\boldsymbol{s}^{\pi}_{r_\mathsf{c}},\boldsymbol{s}_{r_\mathsf{c}}])= t_1+1,r_\mathsf{c}\in \mathcal{R}_\mathsf{c}, |\mathcal{R}_\mathsf{c}|=t_1+1.\label{eq:w3con4}
\end{align}

It is important to note that $w_{\mathsf{H}}(\boldsymbol{s}^{\pi}_{r_\mathsf{c}})$ gives the minimum number of erroneous rows above row $r_\mathsf{c}$ while $w_{\mathsf{H}}(\boldsymbol{s}_{r_\mathsf{c}})$ gives the minimum number of rows affected by row $r_\mathsf{c}$ due to the spreading of errors as a result of the coupling in \eqref{enc:large_w1}. If either $|\mathcal{R}_1|>1$ or $w_{\mathsf{H}}(\boldsymbol{s}_{i,r_1}) >t_1+1$, then the number of affected rows caused by the errors in $\boldsymbol{Y}_{i}$ is strictly larger than $t_1+1$, i.e., $|\mathcal{R}_\mathsf{c}|>t_1+1$, leading to $s_{\min}$ larger than \eqref{eq:app4_contra} and thus is not possible. Note that this will also be case if $w_{\mathsf{H}}([\boldsymbol{s}^{\pi}_{r_\mathsf{c}},\boldsymbol{s}_{r_\mathsf{c}}])> t_1+1$, which cannot happen. In addition, if $ |\mathcal{R}_\mathsf{c}|<t_1+1$, one will get
\begin{align}
 &|\mathcal{R}_\mathsf{c}|<t_1+1 \nonumber \\
 \Rightarrow& \mathop{\E}_{r_\mathsf{c} \in \mathcal{R}_\mathsf{c}} \left[w_{\mathsf{H}}([\boldsymbol{s}^{\pi}_{r_\mathsf{c}},\boldsymbol{s}_{r_\mathsf{c}}])\right] \overset{\eqref{eq:51}}{=} \frac{2s_{\min}-w_{\mathsf{H}}(\boldsymbol{s}_{i,r_1})}{|\mathcal{R}_\mathsf{c}|}\overset{\eqref{eq:app4_contra}}{>} t_1+1\label{eq:61}\\
\Rightarrow& w_{\mathsf{H}}([\boldsymbol{s}^{\pi}_{r'_\mathsf{c}},\boldsymbol{s}_{r'_\mathsf{c}}]) > t_1+1,\exists r'_\mathsf{c} \in \mathcal{R}_\mathsf{c}, \label{eq:contra1}
\end{align}
where \eqref{eq:61} follows that the average number of row errors of the stall pattern must be greater than $t_1+1$ if $|\mathcal{R}_\mathsf{c}|<t_1+1$, which leads to \eqref{eq:contra1} that there must exist an erroneous row with index $r'_\mathsf{c}$ such that the number of row errors is larger than $t_1+1$. However, \eqref{eq:contra1} implies that $|\mathcal{R}_\mathsf{c}|>t_1+1$, which is contradictory to $|\mathcal{R}_\mathsf{c}|<t_1+1$. Thus, \eqref{eq:w3con4} must hold.

Based on \eqref{eq:wL_con0}-\eqref{eq:wL_con1} in Appendix \ref{app3}, \eqref{eq:w3con1_new} guarantees that
\begin{align}
&\left[[\boldsymbol{S}^{\pi}_{i-l+z,l}]_{l=1}^{w-1},\boldsymbol{S}_{i+z}\right] = \boldsymbol{0}, \forall z\in [w-1]\cap (2\mathbb{N}-1) \label{eq:s_con_odd_0}\\
 \Rightarrow &\boldsymbol{S}^{\pi}_{i,l} = \boldsymbol{0}, \forall l\in [w-1]\cap (2\mathbb{N}-1)\label{eq:s_con_odd}.
\end{align}
If \eqref{eq:s_con_odd} is not satisfied, then there will be at least one erroneous row in $[\boldsymbol{Y}^{\pi}_{\Sigma},\boldsymbol{Y}_{\Sigma}]$ defined in Appendix \ref{app3} with at least $t_2+1$ errors, leading to $s_{\min}$ larger than \eqref{eq:app4_contra}, which cannot happen.

We then determine the number of errors of each affected row caused by the errors spreading from the erroneous row in $\boldsymbol{Y}_i$. To get the position of each affected row index, we list all elements of $\mathcal{R}_\mathsf{c}$ as a sequence in ascending order and define a bijective function $g:\mathcal{R}_\mathsf{c} \rightarrow [t_1+1]$ that maps index $r_\mathsf{c}$ to its position of the sequence, i.e., $g(\min\{\mathcal{R}_\mathsf{c}\})=1, g(\min\{\mathcal{R}_\mathsf{c} \setminus \min\{\mathcal{R}_\mathsf{c}\} \})=2,\ldots,g(\max\{\mathcal{R}_\mathsf{c}\})=t_1+1$. Consider that the $g(r_\mathsf{c})$-th affected row is in $\left[[\boldsymbol{S}^{\pi}_{i-l+\bar{z},l}]_{l=1}^{w-1},\boldsymbol{S}_{i+\bar{z}}\right]$, where $\bar{z}=\lceil\frac{q  r_\mathsf{c}}{m} \rceil$ and $\bar{z} \in [w-1]\cap (2\mathbb{N})$ by \eqref{eq:s_con_odd}. Then, $\forall r_\mathsf{c} \in \mathcal{R}_\mathsf{c}$, the following must hold
\begin{align}
\left(w_{\mathsf{H}}(\boldsymbol{s}^{\pi}_{r_\mathsf{c}}),w_{\mathsf{H}}(\boldsymbol{s}_{r_\mathsf{c}})\right) = \big(&  g(r_\mathsf{c}),t_1+1-g(r_\mathsf{c}) \big),\label{eq:rc1}\\
\sum\nolimits_{j=1}^{\frac{m}{q}}w_{\mathsf{H}}(\boldsymbol{s}^{\pi}_{i+\bar{z},l',j})=1,&\forall l'\in \mathcal{L}_{i+\bar{z}} \subseteq [w-1]\cap (2\mathbb{N}), \nonumber \\
&|\mathcal{L}_{i+\bar{z}}| = t_1+1-g(r_\mathsf{c}). \label{eq:rc1_con2}
\end{align}

For \eqref{eq:rc1}, since there are $g(r_\mathsf{c})$ erroneous rows (including the erroneous row in $\boldsymbol{Y}_i$) above the $g(r_\mathsf{c})$-th affected row, hence $w_{\mathsf{H}}(\boldsymbol{s}^{\pi}_{r_\mathsf{c}})\leq g(r_\mathsf{c})$. However, if $w_{\mathsf{H}}(\boldsymbol{s}^{\pi}_{r_\mathsf{c}})< g(r_\mathsf{c})$, then $w_{\mathsf{H}}(\boldsymbol{s}_{r_\mathsf{c}})>t_1+1-g(r_\mathsf{c})$, leading to $|\mathcal{R}_\mathsf{c}|>t_1+1$, which is contradictory to \eqref{eq:w3con4}. Thus, \eqref{eq:rc1} must hold.

As for \eqref{eq:rc1_con2}, it means that the $l'$-th sub-block of $\boldsymbol{Y}^{\pi}_{i+\bar{z}}=[\boldsymbol{Y}^{\pi}_{i+\bar{z},1},\ldots,\boldsymbol{Y}^{\pi}_{i+\bar{z},w-1}]$, i.e., $\boldsymbol{Y}^{\pi}_{i+\bar{z},l'}$, must have only one error, where $l'$ is even according to \eqref{eq:s_con_odd}, $\mathcal{L}_{i+\bar{z}}$ is the collection of these even indices, and $\boldsymbol{s}^{\pi}_{i+\bar{z},l',j}$ denotes the $j$-th row of the corresponding stall pattern matrix $\boldsymbol{S}^{\pi}_{i+\bar{z},l'}$. If $\sum_{j=1}^{\frac{m}{q}}w_{\mathsf{H}}(\boldsymbol{s}^{\pi}_{i+\bar{z},l',j})>1$, then $\boldsymbol{S}^{\pi}_{i+\bar{z},l'}$ must have either more than one non-zero rows or only one non-zero row but with more than one errors. In the former case, there will be at least two erroneous rows in $\boldsymbol{Y}_{i+\bar{z}+l'}$ because $ \boldsymbol{S}^{\pi}_{i+\bar{z},l'}\neq \boldsymbol{0}\Rightarrow \left[[\boldsymbol{S}^{\pi}_{i+\bar{z}+l'-l,l}]_{l=1}^{w-1},\boldsymbol{S}_{i+\bar{z}+l'}\right] \neq \boldsymbol{0}$.
We denote the indices of these two erroneous rows by $r'_\mathsf{c}$ and $r''_\mathsf{c}$, respectively, and $r''_\mathsf{c}>r'_\mathsf{c}> r_\mathsf{c}$. By $\eqref{eq:rc1}$, we know that there are $g(r'_\mathsf{c})-1$ erroneous rows above these two erroneous rows while $\max\{t_1+1-g(r'_\mathsf{c}),t_1+1-g(r''_\mathsf{c})\}$ rows in $\boldsymbol{Y}_{i+\bar{z}+l'+1},\ldots,\boldsymbol{Y}_{i+w-1}$ will be affected. In this case, the total number of affected rows in $[\boldsymbol{Y}^{\pi}_{\Sigma},\boldsymbol{Y}_{\Sigma}]$ will become $t_1+2$, which is contradictory to \eqref{eq:w3con4}. In the latter case, it means that there will be at least two erroneous rows in $\boldsymbol{Y}_{i+\bar{z}}$ because of the transformation of \eqref{eq:int1}. Then, one will arrive at the conclusion that the total number of affected rows in $[\boldsymbol{Y}^{\pi}_{\Sigma},\boldsymbol{Y}_{\Sigma}]$ will become $t_1+2$, which cannot happen. Thus, \eqref{eq:rc1_con2} must hold.

Finally, by applying the arguments of \eqref{eq:w3con1_new}, \eqref{eq:s_con_odd}-\eqref{eq:rc1_con2} to the erroneous row in $\boldsymbol{Y}_i$, we get
\begin{align}
\sum\nolimits_{j=1}^{\frac{m}{q}}&w_{\mathsf{H}}(\boldsymbol{s}^{\pi}_{i,l',j})=1, \nonumber \\
&\forall l'\in \mathcal{L}_1 \subseteq[w-1]\cap (2\mathbb{N}), | \mathcal{L}_1| = t_1+1, \label{eq62a}\\
\Rightarrow \sum\nolimits_{j=1}^{\frac{m}{q}}&w_{\mathsf{H}}(\boldsymbol{s}_{i,l'',j})=1, \nonumber \\
&\forall l'' \in \mathcal{L}'_1 \subseteq [q], | \mathcal{L}'_1| = t_1+1,\label{eq62}
\end{align}
where $\boldsymbol{s}^{\pi}_{i,l',j}$ and $\boldsymbol{s}^{\pi}_{i,l'',j}$ denote the $j$-th rows of stall pattern matrices $\boldsymbol{S}^{\pi}_{i,l'}$ and $\boldsymbol{S}_{i,l''}$, respectively, and $\mathcal{L}_1$ and $\mathcal{L}'_1$ are the corresponding sets of sub-block indices, respectively, with which the sub-block has one error. However, \eqref{eq62a} requires that $\big|[w-1]\cap (2\mathbb{N})\big| \geq t_1+1 \Rightarrow w-1 \geq 2( t_1+1)$ while \eqref{eq62} requires that $q \geq t_1+1$, which are in contradiction with the conditions in Lemma \ref{lem:stall_w3}. Hence, $s_{\min}$ is strictly larger than \eqref{eq:app4_contra}. The proof for the case of $t_2 < t_1$ follows similarly.

\bibliographystyle{IEEEtran}
\bibliography{MinQiu}

\end{document}